\documentclass[journal,twocolumn,10pt]{IEEEtran}

\usepackage{enumitem}
\usepackage{bm}
\usepackage{multirow}

\usepackage{amsmath}
\usepackage{cases}

\usepackage{balance}

\usepackage[]{fixltx2e}
\usepackage[bottom]{footmisc}
\usepackage{xcolor}
\usepackage{soul}
\usepackage{graphicx}
\usepackage{subcaption}
\usepackage{booktabs}
\usepackage[export]{adjustbox}

\usepackage{algorithm}
\usepackage{algorithmic}
\usepackage{caption}
\usepackage{newfloat}
\usepackage{listings}
\usepackage[most]{tcolorbox}

\usepackage{url}

\AtBeginDocument{%
  \providecommand\BibTeX{{%
    \normalfont B\kern-0.5em{\scshape i\kern-0.25em b}\kern-0.8em\TeX}}}

\usepackage{amsmath,amsfonts,amssymb}

\tcbset{
  promptbox/.style={
    colback=gray!10,
    colframe=black!50,
    boxrule=0.5pt,
    arc=3pt,
    left=5pt,
    right=5pt,
    top=5pt,
    bottom=5pt,
    fontupper=\ttfamily\small,
    breakable,
  }
}

\floatstyle{ruled}
\newfloat{listing}{tb}{lst}{}
\floatname{listing}{Listing}

\begin{document}

\title{Debiasing International Attitudes: LLM Agents for Simulating US-China Perception Changes}

\author{
  \IEEEauthorblockN{Nicholas Sukiennik\textsuperscript{1,2}, 
                    Yichuan Xu\textsuperscript{3}, 
                    Yuqing Kan\textsuperscript{3}, 
                    Jinghua Piao\textsuperscript{1,2}, 
                    Yuwei Yan\textsuperscript{4}, 
                    Chen Gao\textsuperscript{2}, 
                    Yong Li\textsuperscript{1,2}} \\
  \IEEEauthorblockA{\textsuperscript{1} Department of Electronic Engineering, Tsinghua University\\}
   \IEEEauthorblockA{\textsuperscript{2} BNRist, Tsinghua University, China}\\
  \IEEEauthorblockA{\textsuperscript{3} Department of Foreign Languages and Literatures, Tsinghua University, China}\\
  \IEEEauthorblockA{\textsuperscript{4} The Hong Kong University of Science and Technology (Guangzhou), China}
}

\maketitle

\begin{abstract}

Large Language Models (LLMs) offer transformative opportunities to address the longstanding challenge of modeling opinion evolution in computational social science. This study investigates how media influences cross-border attitudes—a key driver of global polarization—by developing an LLM-agent framework to disentangle sources of bias and assess LLMs' capacity for human-like opinion formation in response to external information. We introduce an LLM-agent-based framework that models U.S. citizens' attitudes toward China from 2005 to 2025. Our approach integrates large-scale news data with social media profiles to initialize agent populations, which then undergo cognitive-aware reflection and opinion updating. We propose three debiasing mechanisms: (1) fact elicitation, extracting neutral events from subjectively framed news; (2) a devil's advocate agent that simulates critical contextualization; and (3) counterfactual exposure to surface inherent model biases. Simulations with two state-of-the-art LLMs (Qwen3-14b and GPT4o) reveal the expected negative attitudinal trend following media exposure. While all three mechanisms mitigate this trend to varying degrees, results indicate that subjective news framing contributes only modestly to negative attitudes, whereas the devil's advocate agent proves most effective overall, suggesting that intermediate analytical steps can produce more human-like agent opinions. Notably, the counterfactual study reveals contradictory findings across models, suggesting region-specific inherent biases tied to models' geographic origins. By advancing understanding of LLM-based opinion formation and debiasing methods, this study contributes to developing more objective models that better align with human cognitive tendencies.

\end{abstract}


\section{Introduction}

Opinion evolution prediction is a long-standing task in simulation and has implications across a vast array of fields, including cognitive science \cite{schwarz2000emotion}, psychology \cite{edwards_theory_1954}, and behavioral science \cite{simon_theories_1966}. The dawn of large language models (LLMs) allows us to model dynamic and complex human opinion trends at a scale never before possible \cite{piao2025agentsociety, yang2025oasis}. LLMs, via their unfathomably large latent semantic embedding space, have been shown to exhibit strong cognitive reasoning abilities on a variety of tasks \cite{momente2025triangulating}, from mathematics \cite{zhang2024mathverse}, to logic \cite{xiao2024logicvista}, to even software engineering \cite{he2025llm}. However, the ability to answer questions through a built-in chain of thought process \cite{wei2022chain, xu2025softcot}, is insufficient to model long-term thought processes such as trends in opinion evolution because of the reliance of such tasks on long-term memory, which is far too long to be able to be stored in a single prompt, despite increasing context token sizes \cite{chen2023longlora, ding2024longrope}. Therefore, several works have designed explicit architectures to capture and simulate more complex, long-term processes adhering to real-world human behaviors \cite{wang2025what}. The first major breakthrough was the use of memory and reflection mechanisms \cite{park2023generative}, both of which have become nearly standard practice in many agent-based simulation scenarios. 

LLM-powered agent-based simulation, for its part, has been employed to simulate a diverse array of large-scale or long-term (or both) cognitive-behavioral phenomena, such as polarization formation \cite{piao2025agentsociety}, emergence of trust behaviors \cite{xie2024can, huang2024trustllm}, collaboration \cite{zhang2023exploringa, wang2024unleashing, lan2023llmbased, guo2024embodied}, theory of mind \cite{li2023camel, strachan2024testing}. Yet another line of works has aimed to simulate real-world, data-driven scenarios such as election outcomes \cite{zhang2024electionsim}, economic stimuli \cite{li2024econagent}, and social media polarization \cite{gao2023s3}. However, among all these prior works, none have aimed at combining cognitive mechanisms with data-driven real-world macro-trends in international perceptions. While prior works have touched upon international dynamics from various perspectives, such as simulating war \cite{hua2024war} and diplomacy \cite{guan2024richelieu}, these works do not consider the cognitive processes of agents representing real humans at the micro scale to form macro trends. 

Our work aims to bridge this gap with a timely scenario: perceptions of US citizens towards the country of China. This subject has increasingly garnered attention in pop culture as well as academia \cite{li2025media} over the past few years, notably due to its influence on global economic and political dynamics. LLMs provide us with a promising opportunity to address this subject via simulation, as doing so can provide insights not only into the way LLMs can be used to model broad, international political perspectives, but also on the way such opinions are formed -- what drives them, and how they are influenced, whether positively or negatively. In this work, we aim to reproduce the trend in US citizen attitudes toward China using a data and cognitive theory-driven LLM-agent-based simulation framework and a set of debiasing strategies. To do so, we first generate a set of thousands of agent profiles, leveraging a large-scale social survey in combination with social media data from X/Twitter to create a representative and feature-abundant agent pool as a representative sample of the US population. We then design a news exposure-reflection mechanism using real-world news data consisting of more than 100,000 articles across 56 publications. Agents are first exposed to a set number of articles per year and are prompted to reflect on what they read using a cognitive dissonance mechanism, allowing them to rationally update their views in light of new information. At the end of each simulated year, agents conduct a survey assessing their overall attitude towards China. 

In light of overwhelming negative agent attitudes, three debiasing strategies are designed to explore the underlying factors in perception formation and to achieve better fidelity to the real world attitude trend. The results of our simulation scenarios on two state-of-the-art LLMs show that with pure, unmolested media exposure, agents' attitudes end up far more negative than those of real humans. Three debiasing methods, in turn, bring the simulated population closer to reality at varying levels. We analyze the simulation results from three perspectives: (1) average favorability rating per simulation year, (2) favorable and unfavorable ratings, and (3) trend reversal capture, all with respect to the ground truth (real-world) data. 

Overall, our work makes the following core contributions: 

\begin{itemize}
    \item We propose a workflow for agent profiles creation from real-world surveys and social media data as a representative sample of a country's population. 
    \item  We propose a simulation framework that uses extensive real-world data and cognitive mechanisms to reproduce a long-term trend of international attitudes. 
    \item We design mechanisms that successfully debias LLM opinions and alleviate the negative influence of news media to achieve fidelity with real-world trends.
\end{itemize}

The findings provide insights to three distinct groups of people: (1) LLM users, to allow them to better understand the potential biasing factors affecting agent opinion formation, 2) social scientists, providing a framework that can be adapted to any international scenario to simulate macro-trends using large-scale agent populations, and 3) policymakers, to be cautious of using LLM-based agent simulation for support in policy creation without first undergoing comprehensive and effective debiasing.

The remainder of this work is organized as follows. Section \ref{sec:related} examines some of the previous works that are most similar to ours and how ours differs. Section \ref{sec:data} introduces the data collected to create robust agent profiles, serve as news exposure for agents, and surveys used as ground truth for the simulation. Section \ref{sec:framework} introduces the simulation framework in detail. Section \ref{sec:exp&res} includes the experiments conducted and their results, including the vanilla simulation scenario and three debiasing methods, across two LLMs. Section \ref{sec:factors} presents analysis of the  fine-grained factors involved in attitudes simulation including domain and demographic analysis. Finally, the paper concludes with a summary and proposition of future works in Section \ref{sec:conclusions}.

\section{Related Work}
\label{sec:related}
\subsection{Opinion Dynamics Modeling}

The study of opinion dynamics is a broad field that starts with understanding the underlying motivations of human thought processes and trends. The core theories in this field have been developed by social scientists in psychology \cite{edwards_theory_1954}, cognitive science \cite{schwarz2000emotion}, and behavioral science \cite{simon_theories_1966}, and include some key theories as selective exposure theory \cite{frey1986recent}, social identity theory \cite{tajfel2004social}, and cognitive dissonance theory \cite{festinger1962cognitive}, among many others. 

Modeling cognition and the way humans form opinions has been a topic of investigation since the early days of AI \cite{ullman1978ai} and has been addressed with probabilistic methods \cite{de2016learning}, conventional agent based methods \cite{monti2020learning}, deep learning  \cite{kim2022measuring,
zuheros2023explainable}
and now LLMs \cite{siemens2022human}. LLM-based works include Chuang et al. \cite{chuang2024simulating} and Yao et al. \cite{yao2025social}, who demonstrate that LLMs can simulate opinion dynamics in agent-based models.
Mondal et al. \cite{mondal2024large} test LLMs for the presence of cognitive dissonant thought processes, showing that their revealed beliefs often differ from their stated answers., whereas Lee et al \cite{lee2025semantic} propose a method to capture belief interconnectedness and predict new belief formation. Mahowald et al. \cite{mahowald2024dissociating} attempt to distinguish LLMs use of language from its abilities in real-world tasks, revealing a difference in formal vs. functional competence, representing a form of dissonance in cognition. 

Many works also address more specific realms of opinion formation: Piao et al. \cite{piao2025emergenceb} examine the tendency for LLMs to experience human-like polarization on several US political issues, while Hu et al. \cite{hu2024generativea} show LLMs tendency to show in-group favoritism and out-group bias in-line with human expectations. 

Our work takes inspiration from these prior works to infuse a population of agents with a cognitive mechanism to discover their ability to reproduce trends in international perceptions.

\subsection{LLM-based Social Simulation}

One of the pioneering works in LLM social simulation is Park et al.'s Social Simulacra \cite{park2023generative} which simulates a small village where agents create relationships and organize events  of their own accord, enabled by a memory-reflection architecture. Other works, such as Wang et al. \cite{wang2025simulating}, leverage social theories including Ajzen's \cite{ajzen1991theory} theory of planned behavior, to provide theoretical foundation for agent decision-making and introduce desire-driven autonomy. Pang et al. \cite{pang2024selfalignment} use social scene simulation for LLM alignment, showing how considering social consequences improves value alignment. Specific applications of LLM social simulations are used for economic prediction \cite{li2024econagent}, social media emotional contagion \cite{gao2023s3} and spatially aware social movements \cite{mou2025ecolang}. 
Recent works also increase scale of LLM agent societies, with OASIS \cite{yang2025oasis} achieving million-agent simulation for recommendation and social media phenomena, and AgentSociety \cite{piao2025agentsociety} which simulates multiple scales of a society from mobility in a city to emergent social behaviors as polarization and misinformation spread. 

Our work is an LLM simulation based on agents representing the US population, who achieve news-induced opinion evolution.

\section{Preliminary Data Collection}
\label{sec:data}

\subsection{Agent Initialization Data}
\label{sec:agent_data}
\begin{figure}[htbp]
  \centering
  \includegraphics[width=0.95\linewidth]{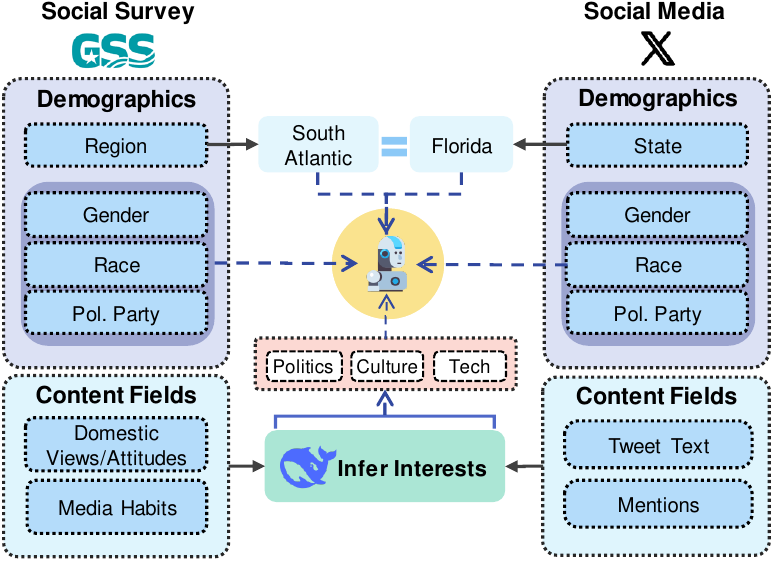}
  \caption{The workflow for generating comprehensive agent profiles based on two real-world datasets.}
  \label{fig:profilegeneration}
\end{figure}

In order to generate a realistic array of agent profiles that can take the form of a representative sample of the US population, and therefore bring about a more realistic simulation, we leverage two public datasets containing real human profiles.
    \textbf{X/Twitter Data\footnote{https://github.com/amazingljy1206/ElectionSim}:} We employ an X/Twitter dataset that contains 3,849 profiles including bio, demographics, and post (Tweet) data from 30 users for each of the 50 US States.
    \textbf{General Social Survey\footnote{https://gss.norc.org/}:} We leverage the 2024 GSS dataset which has 3,309 respondents from all 9 US regions in the and 639 data fields including demographics, preferences on media consumption, and views on various domestic political issues.

The two datasets are processed and merged in the following way, as seen in Figure \ref{fig:profilegeneration}: First the users are matched based on demographics: gender, race, and political party. Because the GSS data assigns users into regions and X data assigns users to states, we reconcile the two by ensuring that the states of the X users fall into the correct GSS regions. However, due to the difficulty of matching all demographic features between both dataset, given that in reality they represent different real-world individuals, we allow a compromise of one demographic feature, so long as it is not region, so that we can maintain a representative US population. In such cases, the gender, race, etc. of the GSS dataset takes precedence. 
Once demographic matching is complete, we use the abundant content fields of both datasets to generate user interests that can be used to inform their news consumption behavior. The X users' post content and the GSS media preferences fields are concatenated together and sent to Deepseek V3 via an API interface, prompting it to apply interests to each user based on a list of 15 topics that correspond to common news categories. After matching the two datasets and applying interests, we are left with approximately 2000 user profiles with 50 features:  8 demographic features, 5 political preferences, 7 media preferences, 30 specific political views on US domestic issues. These features are used in relevant parts of the news exposure and reflection framework described in Section \ref{sec:framework}. The full demographics breakdown of our agent population is provided in the Appendix.

In creating the dataset, we place heavy attention on the correct demographic distributions in order to realize an agent population that is representative of the overall US population. Therefore, we illustrate the demographic closeness between our data and the real US distributions for gender \footnote{https://www.neilsberg.com/insights/united-states-population-by-gender}, political party\footnote{https://www.pewresearch.org/politics/2024/04/09/the-partisanship-and-ideology-of-american-voters/}, region\footnote{https://www.census.gov/library/visualizations/2023/comm \\ /population-change-by-region.html}, and race\footnote{https://www.neilsberg.com/insights/united-states-population-by-race/} in Appendix Figure \ref{fig:a:demographics}. It is observed that our correlations on all demographics are within a fair margin of the real statistics. The exception might be the over-representation of the independent political party in our agent population,
which we attribute to the bias in the GSS data collection. 
Nevertheless, it is still substantially lower than the two mainstream parties, republican and democrat.

\subsection{News Data Collection}

The collection of media content is a prerequisite to modeling the international attitudes evolution, because it is through media, by and large, that people views on such subjects are formed \cite{jost2022cognitive}.
To accomplish this task, we collect news articles about China, over a 20 year period via three methods: API collection via media sources' websites, open source scraping tools, and databases such as Proquest. Our criteria for collecting news is that it should contain the word ``China" or ``Chinese" in the headline or sub-header (if available). Secondly, it must be in the English language, although we do not limit our collection only to US news sources, as outlets from the UK are among the most frequented news website by US citizens \footnote{https://pressgazette.co.uk/media-audience-and-business-data/media\_metrics/most-popular-websites-news-us-monthly-3}. Secondly, we placed heavy emphasis on mainstream news outlets, such as the Guardian, BBC, Financial Times, the Wall Street Journal, etc. Meanwhile, we are sure to include local news sources as well, i.e. those that are not well known countrywide but only in a respective city or state, given that Americans report to consult such sources regularly, despite such consumption declining over time \cite{martin2019local}. The news sources breakdown is provided in the Appendix.
In total, we collect over 100,000 articles from across 68 news publications, with over 90\% of them being from mainstream US or UK news sources, and the rest from local US sources. The news articles cover categories including economics, politics, science, technology, etc., all as it pertains to China. The distribution of articles on a yearly basis and a category basis are provided in Appendix Figure \ref{fig:newsdata}.

\subsection{Ground Truth Attitudes}

The ground truth attitudes of US citizens' perspectives towards China are obtained from two sources: 

\begin{itemize}[leftmargin=*]
    \item \textbf{Pew Research Institute \cite{clancy2025negativea}: } This report from a non-partisan, non-profit US research institution covers Americans' views on China yearly from 2005 until 2025.
    \item \textbf{Gallup. Inc \cite{2023recordlow}:} Gallup is a US-based analytics and polling company, and their survey solicits Americans attitudes towards China at varying intervals between 1979 and 2023. 
\end{itemize}

While the scope and findings of these reports are similar, we opt to use the Pew Research Institute report as the ground truth for the prediction conducted in this paper, given that it reliably covers each year from 2005 until the current year, 2025. In the downstream analysis Section \ref{sec:factors}, Gallup data is used, as they provide more granular results as to the demographic breakdown of responses. Given that the overall trend between the two surveys over the 20 year period is similar, we contend that it is valid to use a different survey for the demographic distribution vs. the overall results. 

\section{Simulation Framework}

\begin{figure*}[htbp]
  \centering
  \includegraphics[width=0.98\linewidth]{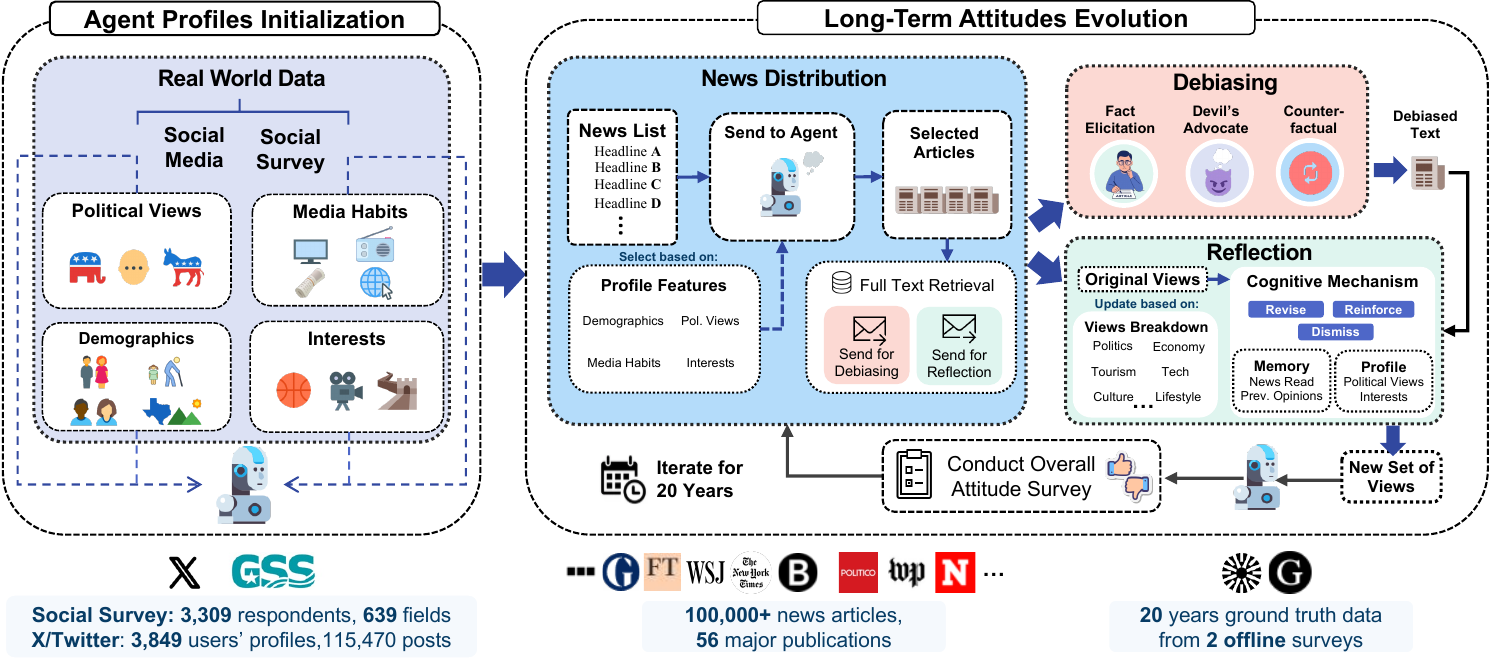}
  \caption{Our framework for macro-scale cognitive-based attitudes evolution and debiasing.}
  \label{fig:framework}
\end{figure*}

\label{sec:framework}

Our simulation framework, seen in Figure \ref{fig:framework}, consists of two major components: agent profiles initialization and long-term attitudes evolution,  The first uses the US demographics distribution data collected and prepared to initialize our agents' profiles. The features, data distribution, and processing methods, are discussed in detail in the previous section, Section \ref{sec:data}. Therefore, in this section we will focus on the second major component: Long-term Attitudes Evolution, which in turn is composed of three parts: (1) news distribution, (2) debiasing, and (3) a reflection mechanism. The three components go through an iterative cycle for each year of the simulation, where, prior to each new iteration, all agents conduct an overall attitudes survey to collect their new perceptions of China after the year of news consumption and reflection. 

\subsection{News Distribution}

The news distribution stage is managed by a distributor module that, for each year, gathers random samples of articles published that year, corresponding to the number of agents in the simulation, and distributes them to each agent. The distributor at first only broadcasts the headlines of each article in the sample to the agents. Each agent sends back a number of article headlines that they would like to read, selecting based on their individual profiles: demographic features, political views, specific interests, and media habits. The distributor then retrieves the full text for the selected articles and sends it back to the agents. The process of sampling is formalized using the two formulas below:
\begin{equation}
\label{eqn:newsdist}
\mathcal{H} = \mathcal{D}_Y(\mathcal{S}(U, \mathcal{A}_Y)),
\end{equation}

\noindent where $\mathcal{H}$ denotes the set of headlines broadcast to agents. $\mathcal{D}_Y$ denotes the distributor module for year $Y$. $\mathcal{S}(U, \mathcal{A}_Y)$ signifies the function that samples $U$ articles from $\mathcal{A}_Y$. $\mathcal{A}_Y$ represents the set of all articles published in year $Y$, and $U$ represents the number of random samples, which is equal to the number of agents. 

In turn, news selection is formulated as
$\mathcal{F}_i = \mathcal{T}(\mathcal{H}_i)$,
where  $\mathcal{F}_i$ signifies the set of full text articles received by agent $i$, which is determined by $\mathcal{H}_i$, the set of headlines selected by agent $i$, where $H < \mathcal{A}_Y$. $\mathcal{H}$ represents the full text retrieval function. 

Some considerations are involved in this design. First of all, the headlines broadcast to each agent do not contain any other information, such as publication etc. The reason for this is to prevent bias based on strong correlations between political affiliations and certain media outlets. In other words, the agent will leverage his/her profile to determine interest in the article solely based on the headline. We also opt not to use any form of targeted recommendation to match agents with articles. This is because, rather than trying to simulate a user scrolling a social media feed and clicking on an article of interest, which entails bias presented by the platform \cite{ribeiro2018media}, we want our simulation to simulate the way people obtain news across platforms and different forms of media in their day to day life, given that not all media a person consumes is obtained via recommendations.

\subsection{Debiasing}

\begin{figure}
  \centering
  \includegraphics[width=0.99\linewidth]{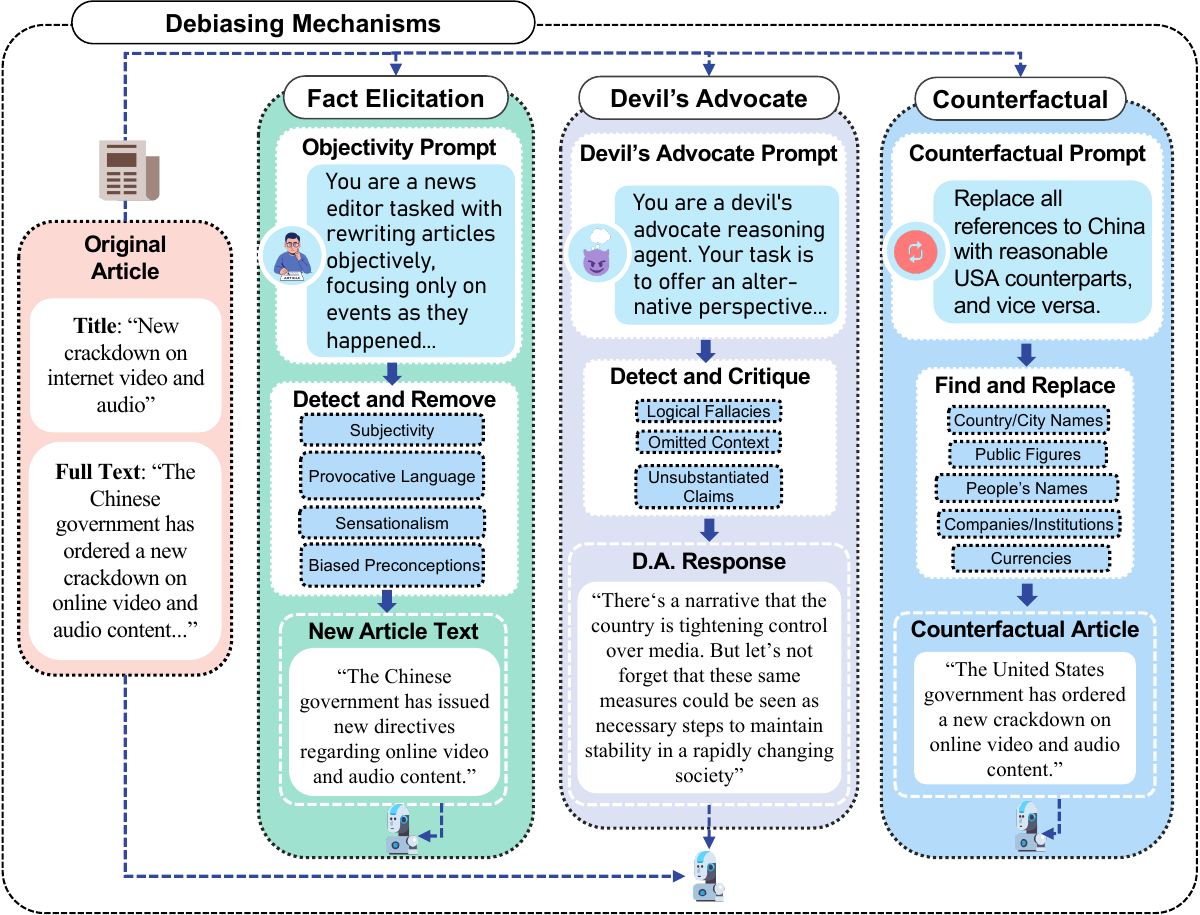}
  \caption{Intervention mechanisms for news exposure.}
  \label{fig:interventions}
\end{figure}

Once articles are selected they are sent to the next step which is either reflection (un-debiased) or one of three debiasing stages. The three debiasing methods are illustrated in Figure \ref{fig:interventions}. Formally, debiasing is defined as:

\begin{equation}
\label{eqn:debiasing}
\mathcal{F}_i^\text{debiased} = \mathcal{D}\left(\mathcal{F}_i\right)
\end{equation}

\noindent where $\mathcal{F}_i$ is the set of raw (non-debiased) full-text articles received by agent $i$ (as defined in Equation \ref{eqn:newsdist}), $\mathcal{D}(\cdot)$ denotes the debiasing function that transforms raw articles into debiased articles, and $\mathcal{F}_i^\text{debiased}$ is the set of debiased articles sent to agent $i$ (the final output before delivery).

Below, we describe the three types of debiasing in detail: fact elicitation, devil's advocate, and counterfactual debiasing.

\textbf{Fact Elicitation (FA).} The fact elicitation mechanism is based on the premise that people's opinions may be affected by the emotional provocation, or sensationalism, of news articles \cite{brown2018new}, thereby causing them to form exaggerated opinions of a certain subject by processing it emotionally rather than logically.

The fact elicitation mechanism works as follows. First, the selected news articles are sent to an ``editor" agent, with a prompt explaining how articles should be revised, removing subjective or provocative language or other bias indicators, focusing primarily on conveying objective facts, i.e. the events as they happened. The revised news article, now a sequence factual events, is then sent to the agent in place of the original article. The rest of the simulation process remains the same.

In order to validate that our fact elicited method reliably removes subjectivity-related bias from the original news articles, we conduct a real-world survey, the results of which are presented in the Appendix.

\textbf{Devil's Advocate (DA).} The devil's advocate agent is intended to simulate a human reasoning process that is critical or questioning of a given narrative, via a process called analytical thinking \cite{pennycook2019lazy}, rather than taking said narratives at face value. To simulate this process, news articles are first sent to a DA agent. The DA agent is given the task of critiquing potential weak points in reporting, such as logical fallacies, omitted context, or unsubstantiated claims, with the end goal of providing an alternative perspective on the events covered in the news media. 
The DA agent's response is then sent to the citizen agent alongside the article's original text, therefore allowing it to proceed to reflection having been exposed to the news content with the context from the extended reasoning DA response. 
Because in this case the original article itself is not altered, but sent with supplementary information, Equation \ref{eqn:debiasing} is modified to become: 

\begin{equation}
\mathcal{F}_i^\text{final} = \mathcal{F}_i \cup \mathcal{R}_i^\text{DA}
\end{equation}

\noindent where $\mathcal{F}_i$ denotes the original, unmodified set of full-text articles received by agent $i$, $\mathcal{R}_i^\text{DA}$ represents the DA agent’s extended reasoning response for agent $i$, and $\mathcal{F}_i^\text{final}$ is the combined set of original article content and DA agent context sent to agent $i$ for reflection.

\textbf{Counterfactual (CF).} The last debiasing method entails swapping all references to the original subject of the attitude survey with another subject. In this case, the subject is China, and the agents are ``US citizens"; therefore all China related references are swapped for US counterparts, and vice-versa. In line with behavioral theory on ingroup favoritism \cite{brewer1999psychology} we expect that agents will rate their own country more favorably than they would if the subject were China. This method also allows us to test for implicit bias in the LLMs themselves, given that two models are tested: one developed in the US, and other in China.  

\subsection{Reflection Mechanism}
After the news media is processed by the debiasing mechanisms they are sent to the agents for reflection. For the control (non-debiased) experiment, the original news content is sent.  Reflection takes place on a specified batch of articles at a time (equal to or less than the number of articles selected). 

Reflection on a batch articles involves three steps: a cognitive mechanism based on cognitive dissonance theory \cite{festinger1954theory}, an opinion decomposition stage, where the information synthesized and new cognitions are discretized under a set of topics on which agents will update views.

 \textbf{Cognitive Mechanism. } The cognitive mechanism is designed a way for an agent to rationally handle new information according to whether it aligns with or contradicts existing beliefs. First the agent is prompted to decide whether or not there are any contradictions. If there are none, then the given cognition will not change. But if there are contradictions, then the agent can choose from among three option: revise, reinforce, or dismiss. Revise means that the agent will essentially accept that their prior view was incomplete or incorrect, and adopt a new belief consistent with the new information. Reinforce means that the agent refuses to let go of her existing opinion, and in order to do so she adds a new cognition to bridge the gap between the existing and the old one, essentially rationalizing the existing opinion in the face of contradictory information. Finally, the last option is dismiss, where the agent opts to reduce the importance of the belief entirely, bypassing the need to rationalize it at all. During the cognitive process, the agent will also consult her own profile to help determine which beliefs are strong/important, in essence providing an implicit values framework. 
 
 Agents retain summarized domain-level opinions and prior-year overall attitudes rather than verbatim memories of all previously consumed news. This design choice prevents unbounded context growth while approximating how humans retain abstractions rather than exact textual recall. We leave explicit memory decay or forgetting mechanisms to future work, noting that incorporating such effects may further improve realism for longer time horizons
 \\
 
\textbf{Opinion Decomposition. }
    Once the cognitive process has been undergone, the new cognitions are broken down into a set of categories which reflect the topics covered by the viewed news. The agent will consult his new cognitions and then update the opinions one by one, skipping those not covered by the news and on which no cognitions were formed. The agent is prompted to quantify a valence between -2 and 2 according to whether his perception of the topic has improved or declined (as it pertains to China), where the valences are all initialized to 0. The opinion update procedure is formalized as follows: 
\begin{equation}
o_{i,d}^{y+1} = 
\begin{cases}
o_{i,d}^{y} + \Delta v_{i,d} & \text{if } d \in \mathcal{D}_{\text{news}}, \\
o_{i,d}^{y} & \text{otherwise},
\end{cases}
\end{equation}
where $\Delta v_{i,d} = g(\mathcal{C}_{i,d}) \in [-2, 2]$.
Here $o_{i,d}^{y}$ denotes the opinion of agent $i$ on domain $d$ at year $y$. $o_{i,d}^{y+1}$ denotes the updated opinion after processing news. $\mathcal{D}_{\text{news}}$ denotes the set of domains/topics covered in the viewed news articles. $\Delta v_{i,d}$ denotes the valence change for domain $d$ which is formed by processing the new cognitions $\mathcal{C}_{i,d}$ formed by agent $i$ about domain $d$ after processing the news.

Once the full reflection is complete, the agent will engage in an overall attitude survey with the aim of consolidating his/her evolving views on China over the past year. The survey emulates the format of the real Pew survey on which the simulated results are compared. There is one question in the survey and it is:

\begin{quote}

\textit{On a scale from 0–4, where: \\
1 = Very unfavorable, 
2 = Somewhat unfavorable, \\
3 = Somewhat favorable, \\
4 = Very favorable, \\
0 = Don't Know / Refuse to Answer.\\ }
\textit{How would you rate your current opinion of China? Please respond with a number (1, 2, 3, 4, or 0).}
\end{quote}

To respond to the survey, the agent first loads in their overall attitude from the previous year, as well as the newly updated domain opinions. The agent's profile remains in memory to add a ``human" element to the decision. The domain opinions are then weighted based on exposure (number of times the topic was covered in read news) and then converted to overall valence. The overall valence is calculated as follows: 

\begin{equation}
\bar{o}_i = \frac{\sum_{d \in \mathcal{D}} o_{i,d} \cdot e_{i,d}}{\sum_{d \in \mathcal{D}} e_{i,d}}
\end{equation}

\noindent where $\bar{o}_i$ represents the overall weighted attitude score for agent $i$ toward China. $o_{i,d}$ denotes agent $i$'s opinion valence for domain $d$. $e_{i,d}$ signifies the exposure count for agent $i$ in domain $d$, representing how many times the agent has encountered news about that domain, and $\mathcal{D}$ is the set of all domains.

This new overall valence, the sub-opinions, and the profile are then considered in making the final decision for the survey response. Once the survey is sent back for all users, the simulation moves on to the next iteration (simulation year).

\section{Experiments and Results}
\label{sec:exp&res}

We conduct experiments to test each debiasing strategy in addition to a control (non-debiased) experiment with two state-of-the-art LLMs: GPT4o and Qwen3-14b. Each experiment is conducted with 50 agents randomly initialized from the population of 2000 generated profiles based on real-world data as described in section \ref{sec:agent_data}. The distributor module is configured to send a unique random sample of 50 headlines to each agent from each year's news. Each agent is configured to read and reflect on 5 news articles of their choice per year. As such, 5,000 unique articles are processed per experiment, meaning that in total, we process 20,000 unique article-agent pairings per model. Qwen3-14b and GPT-4o are used for their high performance \cite{yang2025qwen3}\cite{hurst2024gpt}; they are also chosen to represent two regions central to our problem scenario: the U.S. and China. The average number of tokens per news article is 658. Therefore, accounting for the fact that the three debiasing methods involve processing articles twice (once for the debias agent and once for the citizen agent) across the 4 experiments (3 debiasing +1 control), we send approximately 92 million tokens to each model. Both models are prompted via their respective API interfaces. 

Below, we analyze the experimental results on three dimensions: (1) overall attitude scores, (2) attitude trend decomposition, and (3) trend reversal accuracy. 

\subsection{Overall Attitude Scores}
\label{sec:overallattitudes}

\begin{figure*}[tb]
    \centering
    \begin{subfigure}[t]{0.74\textwidth}
        \centering
        \begin{subfigure}[t]{\textwidth}
            \centering
            \includegraphics[width=0.99\textwidth]{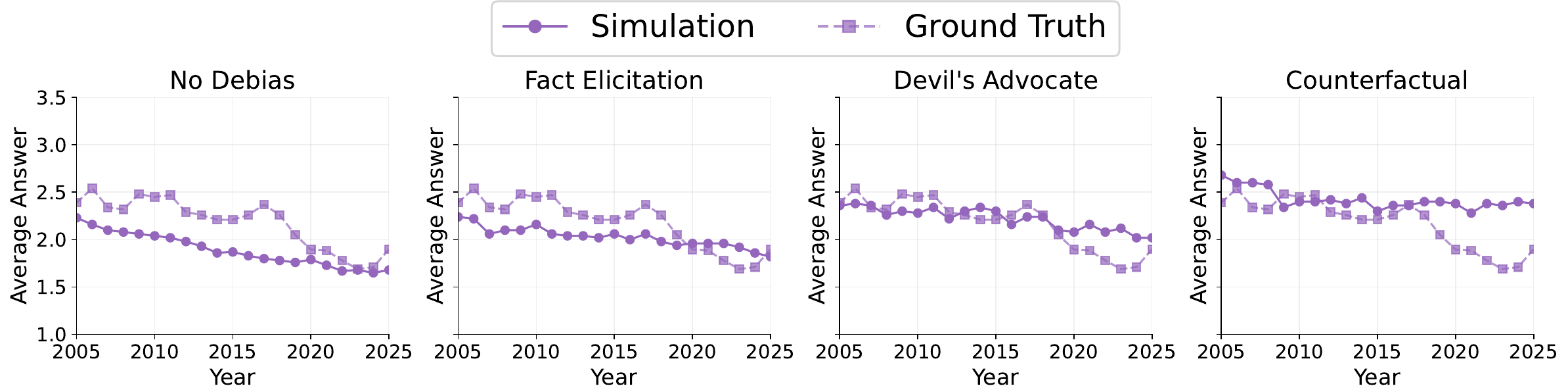}
            \caption{GPT-4o}
            \label{fig:GPT4o_trend}
        \end{subfigure}
        \vspace{1em}
        \begin{subfigure}[t]{\textwidth}
            \centering
            \includegraphics[width=0.99\textwidth]{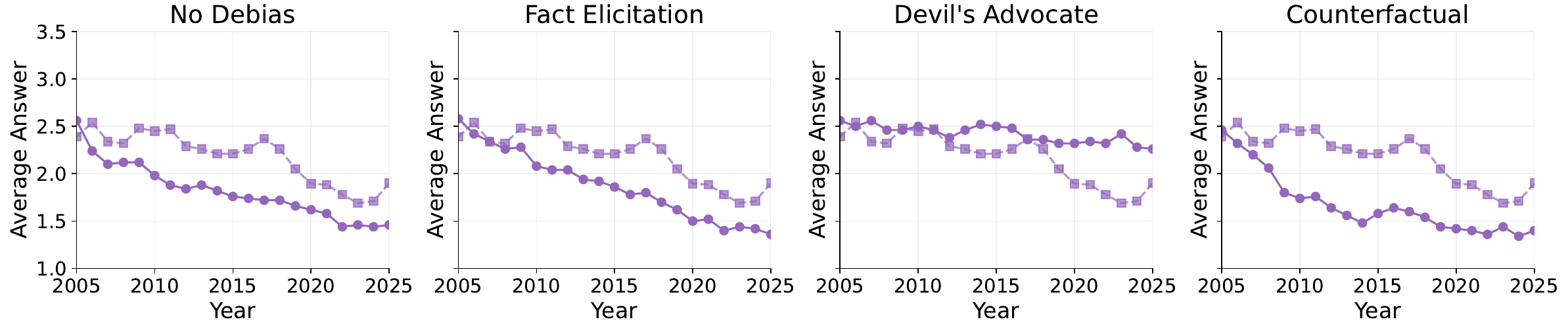}
            \caption{Qwen3}
            \label{fig:Qwen3_trend}
        \end{subfigure}
        \label{fig:average_scores_trend}
    \end{subfigure}
    \hfill
    \begin{subfigure}[t]{0.25\textwidth}
        \centering
        \setlength{\parskip}{0pt} %
        \noindent
        \begin{subfigure}{0.99\linewidth}
            \centering
            \includegraphics[width=\linewidth]{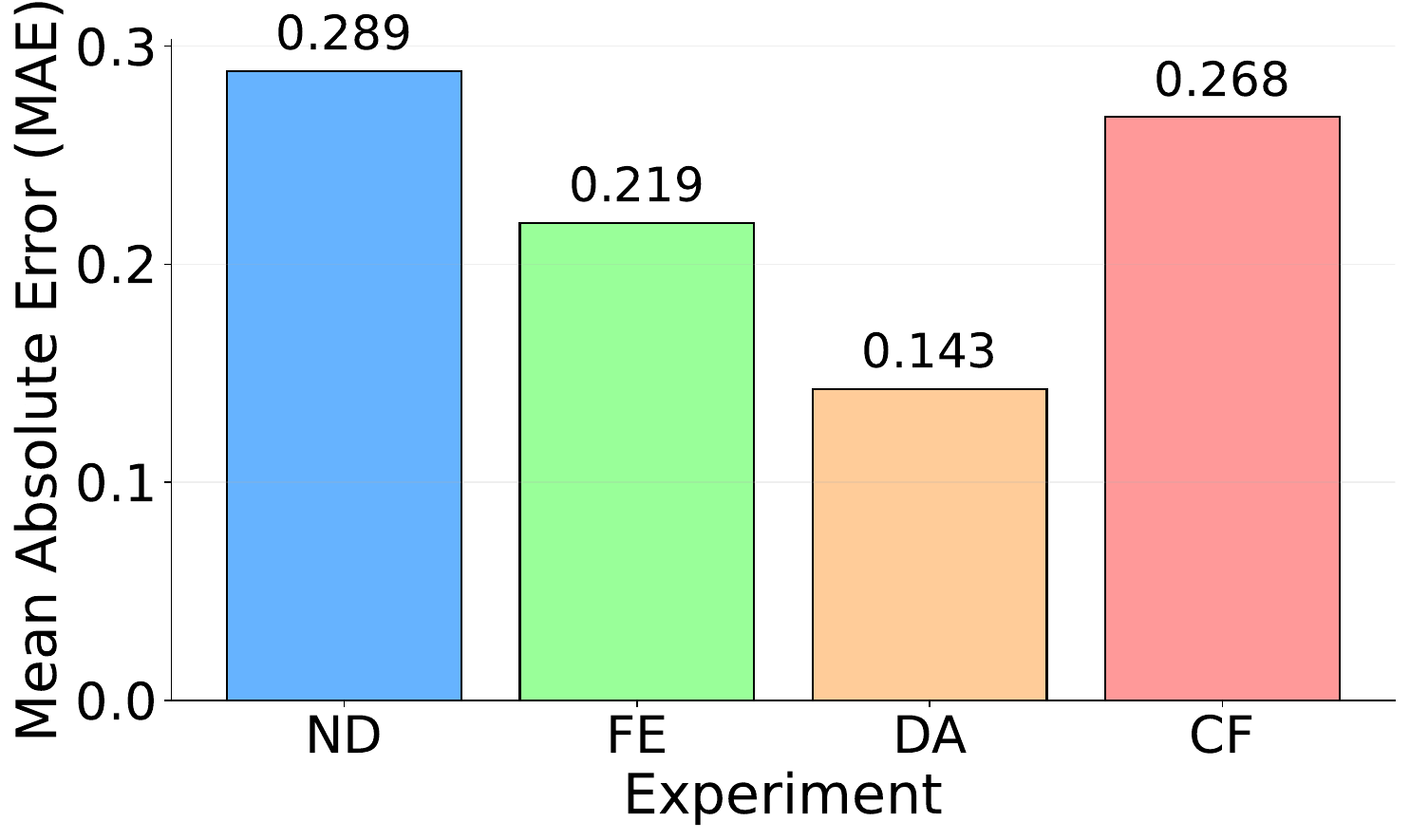}
            \caption{GPT4o}
            \label{fig:GPT4o_mae}
        \end{subfigure}
        \par %
        \vspace*{1em} %
        \begin{subfigure}{0.99\linewidth}
            \centering
            \includegraphics[width=\linewidth]{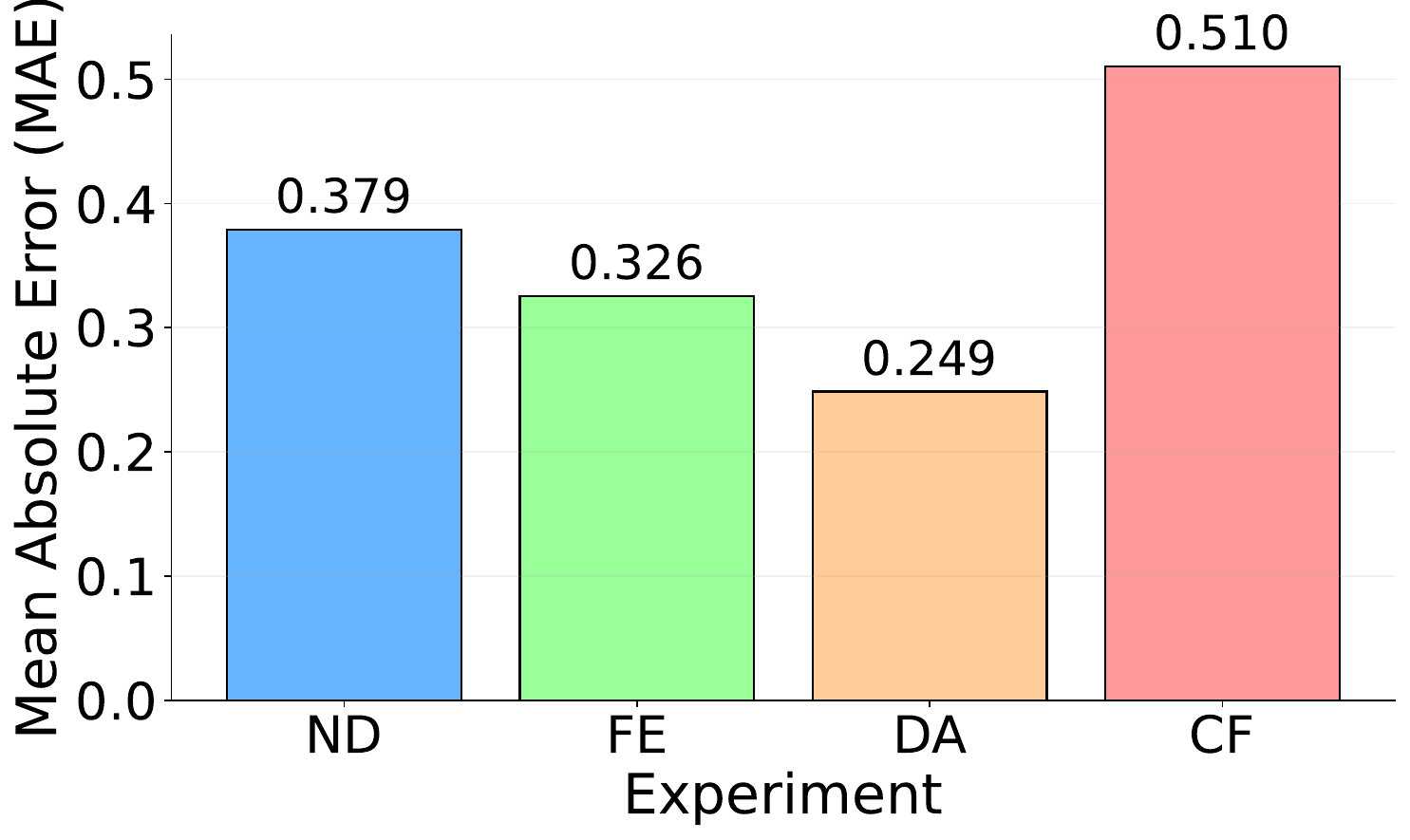}
            \caption{Qwen3}
            \label{fig:Qwen3_mae}
        \end{subfigure}
        \label{fig:average_scores_bars}
    \end{subfigure}

    \caption{Average score trends over time (left) and MAE comparisons (right) for GPT-4o and Qwen3 across four experiment types. }
    \label{fig:combined_avg_scores_trend_mae}
\end{figure*}

The first analysis is conducted on the dimension of overall attitude scores, computed as the average of the attitude survey scores per year for each experiment. This analysis provides insight into the overall characterization of opinion formation among agents for each given debiasing method and the control, revealing which methods align best and worst with the real-world trend.

Figure \ref{fig:combined_avg_scores_trend_mae} (left) shows the average attitude results for each experiment and each model, whereas Figure \ref{fig:combined_avg_scores_trend_mae} (right) quantifies their accuracy with respect to ground truth as measured by mean absolute error (MAE), where the lower, the more accurate. Looking at the GPT-4o qualitative results, we note that the simulation results for all four experiments show a downward trend, largely matching the slope of the real-world trend. However, each experiment causes what could be described as a y-axis shift in the trendline. Looking at the quantitative results, we note that the Devil's Advocate debiasing method results in the most accurate trend with respect to the ground truth, with an MAE of 14.3\%, as opposed to the control, which has the largest error at 28.9\%. For Qwen3-14b, on the other hand, the trendlines vary both in slope and y-axis position according to the experiment. Quantitatively, the Devil's Advocate experiment also shows the best performance at 24.9\% MAE, whereas the counterfactual experiment has the highest error at 51.0\%. We also note that the accuracy w.r.t ground truth is much better for GPT-4o for all experiments. Meanwhile, for both models, the Fact Elicitation method performs second best.

Interpreting these results, we first note that LLM, in the control scenario (no debiasing), outputs responses that are significantly more negative than the real world attitudes over the surveyed period, which could imply the presence of some inherent bias in LLM, or an unrealistic (non-humanlike) way of processing information to inform opinion formation \cite{lin2025six}. This is expected, given that our agents are very simple compared to real humans in that they do not have views or ideas formed over years and years of life experience but rather a set of demographics and views assigned and stored in memory \cite{xiang2023language}. The fact that these assigned views and demographics are not sufficient to create lifelike opinion could reveal a weakness of LLM agents in social simulations, and could point to areas for further investigation with the aim of improving the realism of agents' views. 

As such, we conclude that the reason the Devil's Advocate debiasing method achieves the closest fidelity with the ground truth is because it emulates the cognitive process that a human undergoes when being confronted with new information: a person will cross-check it with prior information they know or believe to be true, and make a conclusion in light of such context -- effectively making sure it crosses a barrier of logic and factual robustness within that person's mental framework of understanding the world, prior accepting it, and therefore rejecting overly biased or sensationalistic rhetoric instead of taking it at face value \cite{ennis1987taxonomy}. 

 Meanwhile, the fact that the least accurate method is different across the two models (No Debias and Counterfactual, for GPT-4o and Qwen3-14b, respectively) causes us to ask a new question: why would counterfactual news exposure increase the error between the simulation results and ground truth for one model but not the other? To answer this question, we explore the results as decomposed into favorable and unfavorable responses in the section below.

\subsection{Attitude Trend Decomposition}
\label{sec:trenddecomposition}
\begin{figure*}[tb]
    \centering
    \begin{subfigure}[b]{0.65\textwidth}
        \centering
        \begin{subfigure}[t]{\textwidth}
            \centering
            \includegraphics[width=0.99\textwidth]{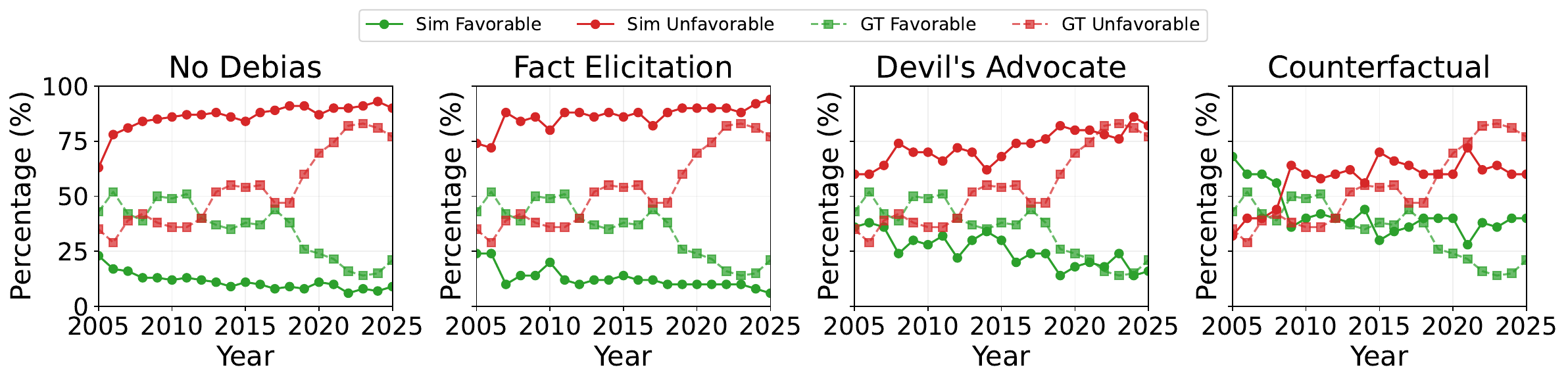}
            \caption{GPT-4o}
            \label{fig:GPT4o}
        \end{subfigure}
        
        \vspace{1em} %
        
        \begin{subfigure}[t]{\textwidth}
            \centering
            \includegraphics[width=0.99\textwidth]{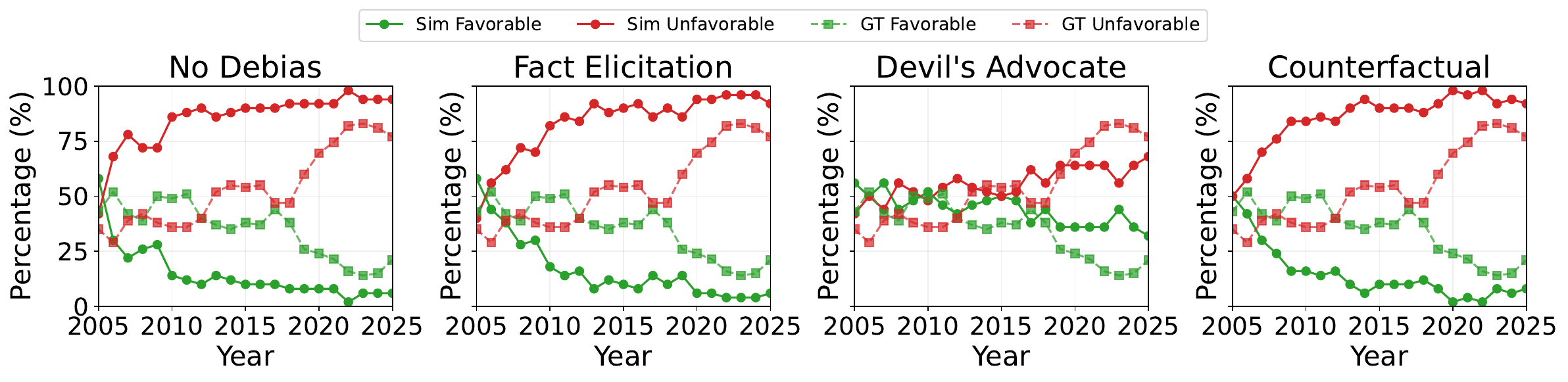}
            \caption{Qwen3}
            \label{fig:Qwen3}
        \end{subfigure}
        \label{fig:decomposed_trends}
    \end{subfigure}
    \hfill %
    \begin{subfigure}[b]{0.34\textwidth}
        \centering
        \setlength{\parskip}{0pt} %
        \noindent
        \begin{subfigure}{0.99\linewidth}
            \centering
            \includegraphics[width=\linewidth]{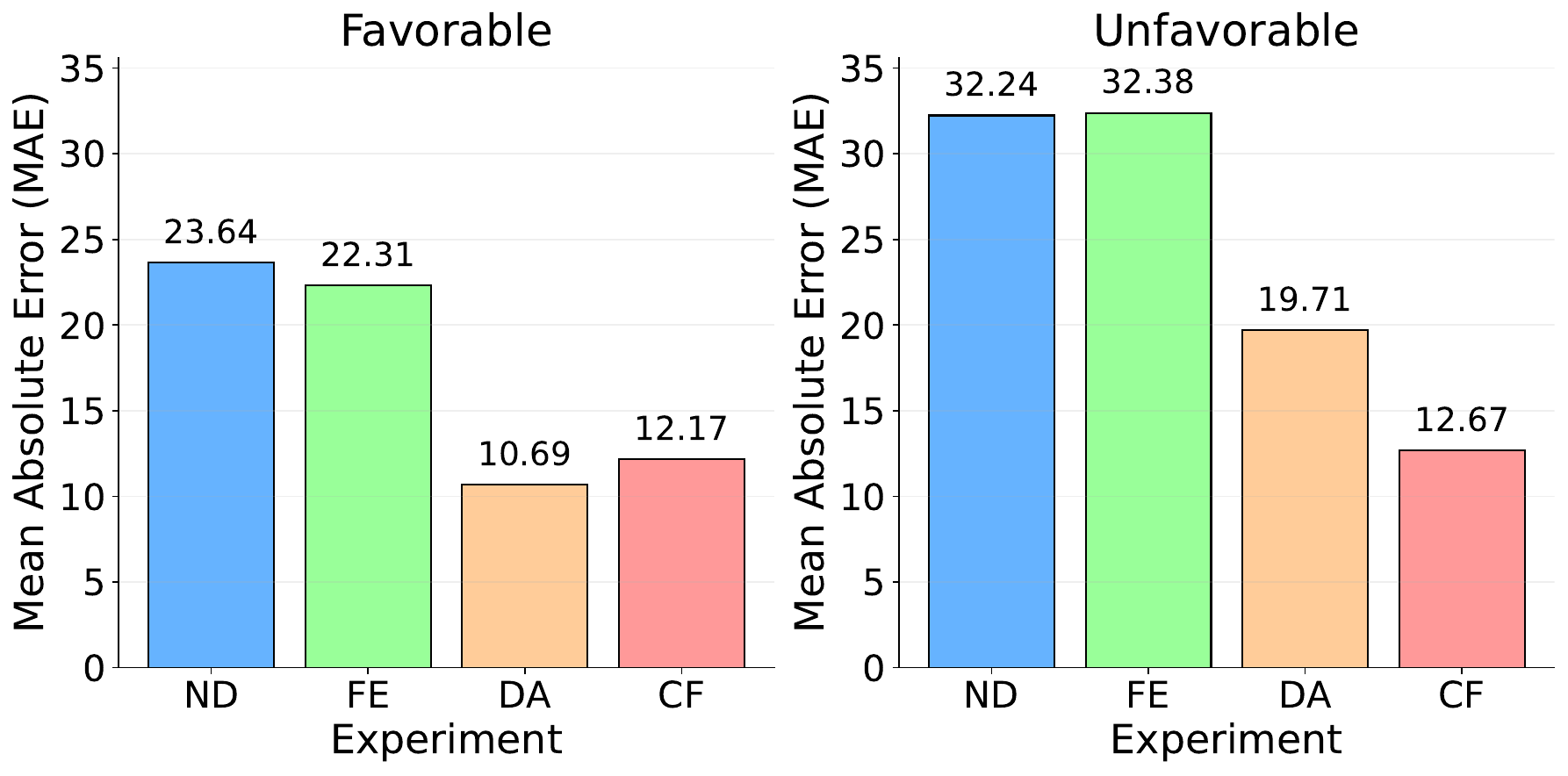}
            \caption{GPT-4o}
        \end{subfigure}
        \par %
        \begin{subfigure}{0.99\linewidth}
            \centering
            \includegraphics[width=\linewidth]{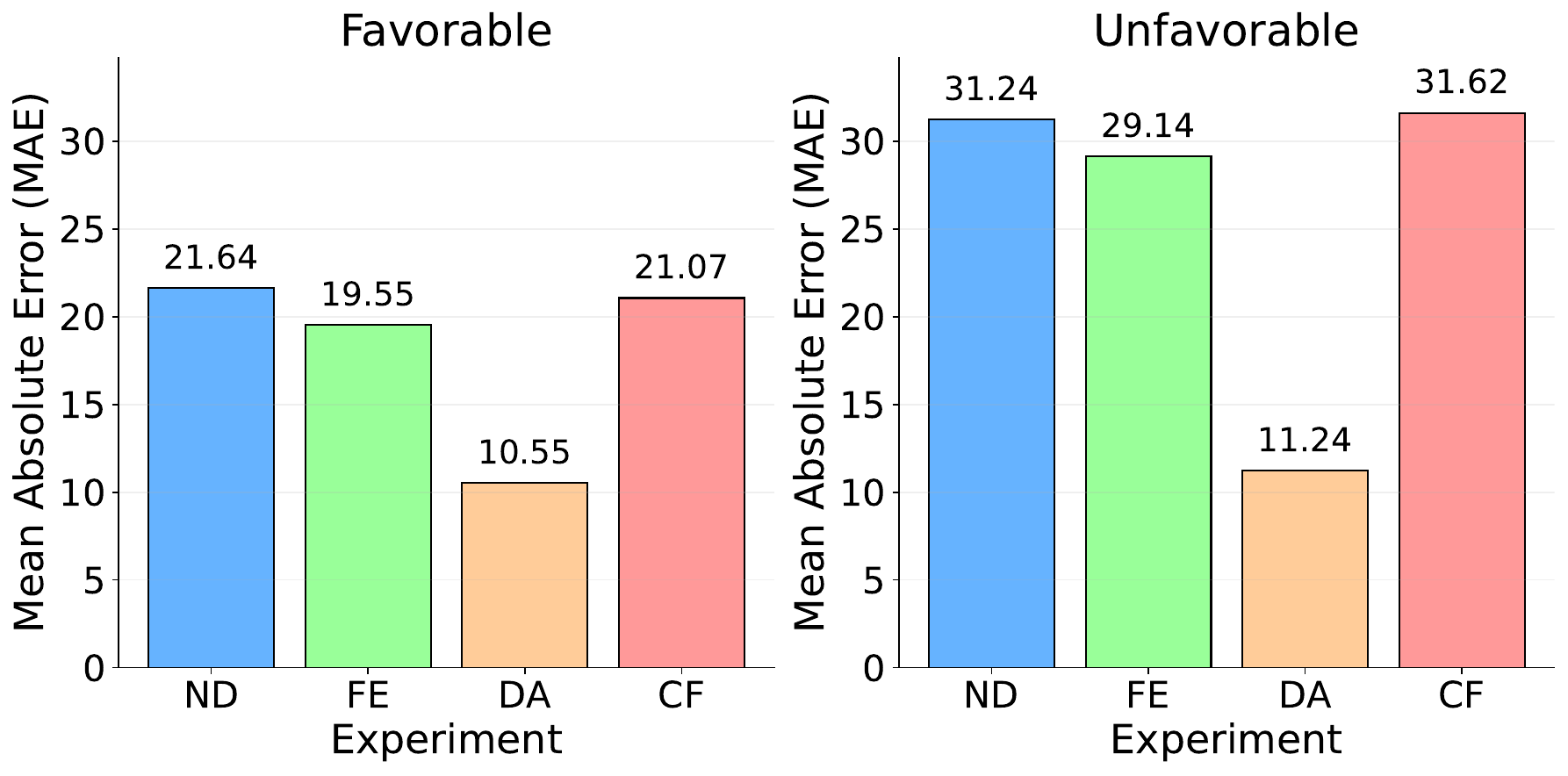}
            \caption{Qwen3}
        \end{subfigure}
        \label{fig:decomposed_bars}
    \end{subfigure}

    \caption{Positive/negative attitude rating trends (left) and MAE comparisons (right) for GPT-4o and Qwen3 across four experiment types}
    \label{fig:combined_decomposed_trends_mae}
\end{figure*}

In this section we want to discover what the drivers are that cause the difference in real-world trend alignment across the two models. To answer this, we illustrate the decomposed trendlines for each experiment: the percentage of favorable and unfavorable responses over the simulation period -- in Figure \ref{fig:combined_decomposed_trends_mae} (left) and then compute the MAE of each of these trendlines with respect to the decomposed ground truth.
The figure shows the percentage of agents who ranked their attitude towards China as favorable (i.e. positive) vs. those that ranked it as unfavorable (negative), along with the corresponding ground truth statistics. A survey response of 3 or 4 is grouped together as favorable, and 1 or 2 is classified as unfavorable.

We also quantify the accuracy for each of the decomposed trendlines for each model in Figure \ref{fig:combined_decomposed_trends_mae} (right). From the decomposed accuracies, we can now pinpoint the underlying drivers of disparity between the two models. Namely, while GPT-4o has strong fidelity to ground truth for both the favorable and unfavorable trendlines for Counterfactual, Qwen3-14b, in contrast, has a wide disparity between the MAE of favorable and unfavorable. More specifically, the counterfactual experiment causes agents attitudes to skew very negatively very quickly, causing an MAE of 31.6\% w.r.t the ground truth trendline for unfavorable, while the favorable MAE for the same model is 21.1\%, falling below that of the No Debias experiment. Meanwhile, GPT-4o, in turn, shows the opposite phenomenon: the MAE for the unfavorable trendline is much higher than that of favorable for the No Debias experiment. In sum, this causes No Debias to have the poorest performance for GPT-4o, whereas Counterfactual has the poorest performance for Qwen3-14b.

Interpreting these results, we suggest that some inherent bias to the models themselves may be the factor that led to such disparity. Specifically, models may have a tendency to look more positively on the news that discusses its own country of origin -- i.e. Qwen will see media more positively if the subject is China, and negatively if the subject is the United States, while, conversely, GPT-4o will see news about China more negatively than the same news stories with U.S. swapped in China's place. Such conclusions echo the discovery by Lu \textit{et. al} that models exhibit cultural tendencies that are more aligned with their country of origin \cite{Lu_Song_Zhang_2025}. 
It is important to emphasize that the counterfactual debiasing method is not intended as a realism-improving intervention, but rather as a diagnostic tool. The observed divergence between GPT-4o and Qwen3-14B reflects a combination of (i) in-group favoritism mechanisms known from social psychology \cite{fu_evolution_2012} and (ii) model-internal cultural priors induced during pretraining. Because these two effects are inseparable in black-box LLMs, we interpret counterfactual exposure as revealing latent bias rather than correcting it, and therefore exclude it from downstream analysis in Section \ref{sec:factors}.

We also note that in our simulations, the favorable and unfavorable trendlines are exact mirror images, whereas for the ground truth, there is slight variation between them. To understand why this is, we consider that real humans may or may not consume news regularly, or otherwise may not consume news about China, and therefore may not have an opinion, and weould therefore select ``Don't Know" in their response to a survey about their opinion on China. On the other hand, our agents have no choice but to view news articles related to China for each simulation year. Therefore, it is intuitive that few agents will choose the ``Don't Know" option as their survey response. This is validated in Appendix Figure \ref{fig:a:dontknow}, which shows that across all 4 experiments, very few agents chose the ``Don't Know" option, albeit there were more ``Don't Know" responses for GPT-4o in the No Debias experiment and at the beginning of the simulation.

\subsection{Trend Reversal Analysis}
\label{sec:reversalanalysis}

In this section, we look at another dimension of the results: trend reversals, and how well each experiment captures them per model. This analysis dimension is a crucial part of understanding the evolution of long-term international attitudes, given that their real-world reversals often coincide with significant diplomatic or political events. Therefore, addressing this dimension captures a complementary element of the simulation results beyond pure error in trend reproduction as treated in the previous sections. 

\begin{figure*}[tb]
    \centering
    \begin{subfigure}{1.0\columnwidth}
        \centering
        \includegraphics[width=\linewidth]{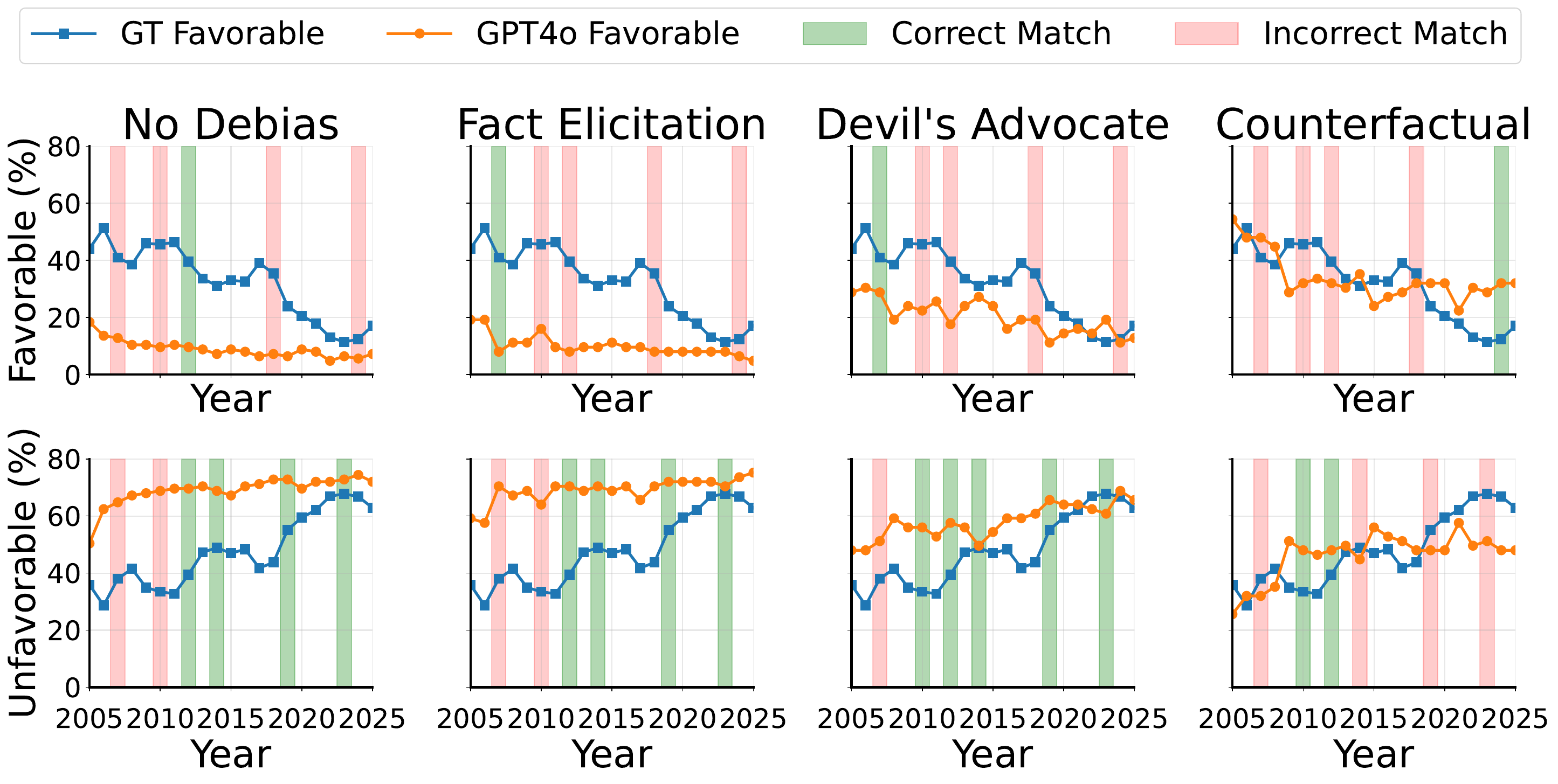}
        \caption{GPT4o}
    \end{subfigure}
    \hfill
    \begin{subfigure}{1.00\columnwidth}
        \centering
        \includegraphics[width=\linewidth]{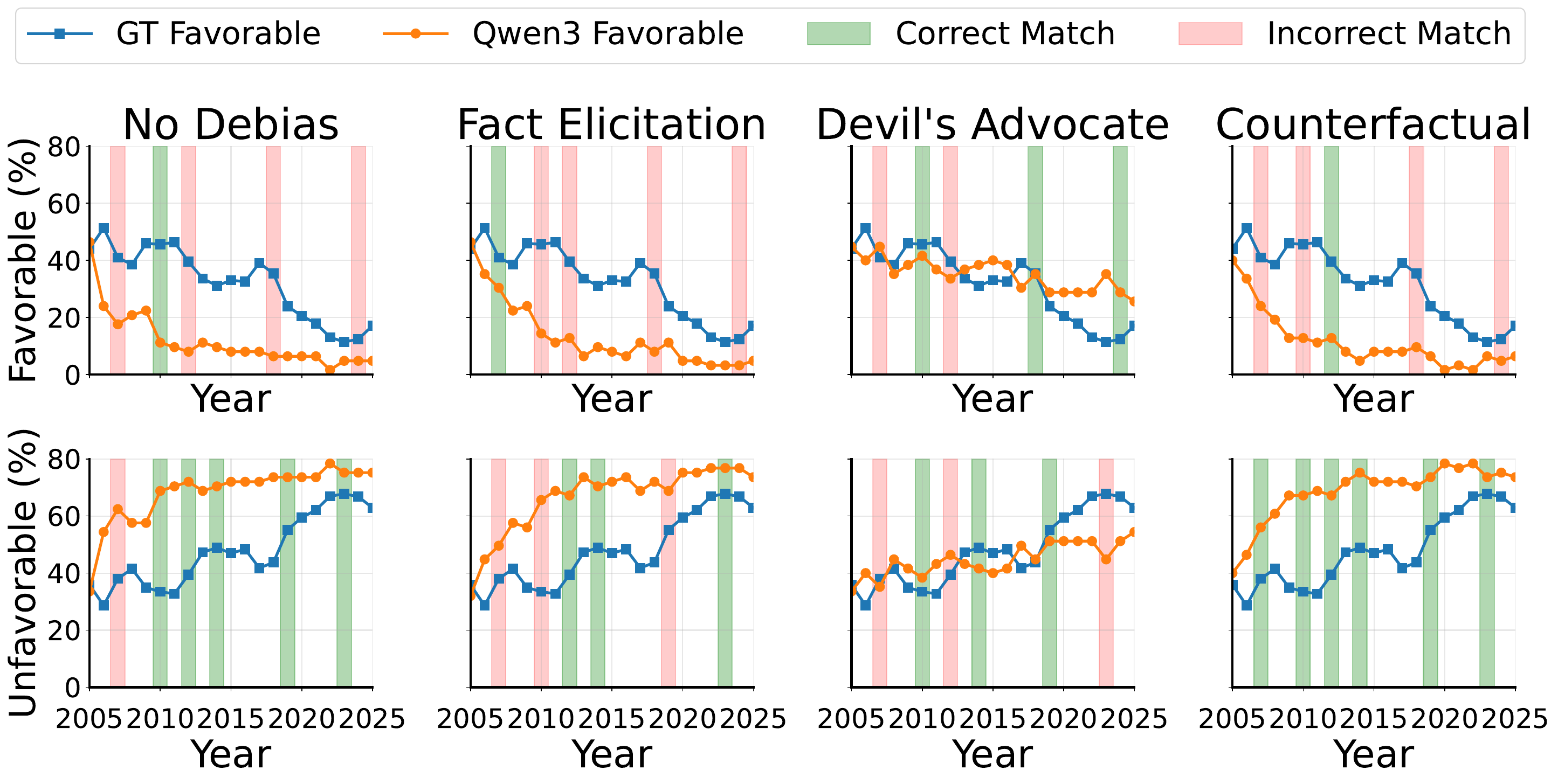}
        \caption{Qwen3}
    \end{subfigure}
    \caption{Visualization of the reversal matching for each of the four experiment types on two models. Reversals are classified as correctly matched if the simulation switches direction within one year of the ground truth switch. }
    \label{fig:reversals_trendlines}
\end{figure*}

Whether a reversal is captured correctly is defined by whether the slope of the simulation trendline reverses direction in the same year as the ground truth trendline (strict) or within one year before or after the ground truth reversal (tolerant). Formally, we define first differences as

\begin{equation}
    \Delta G(y) = G(y) - G(y-1)
\end{equation}

\noindent for ground truth and 

\begin{equation}
  \Delta S(y) = S(y) - S(y-1)
\end{equation}

\noindent for simulation, where $y$ denotes the year index. A reversal at year (y) is said to occur when the slope changes sign, i.e., 

\begin{equation}
        \Delta G(y) \Delta G(y-1) < 0
\end{equation}
\noindent for the ground truth, and analogously for the simulation.

Let $(y^*)$ denote a year in which a ground-truth reversal occurs. A simulated reversal is considered correctly matched if there exists a year $y \in [y^*-\tau, y^*+\tau]$, where $\tau \in \{0,1\}$ denotes the temporal tolerance, such that both of the following conditions hold:

\begin{equation}
\Delta S(y)\Delta S(y-1) < 0,
\end{equation}
\noindent  indicating that the simulation exhibits a reversal within the tolerance window, and 
\begin{equation}
\operatorname{sign}\bigl(\Delta S(y)\bigr)
=
\operatorname{sign}\bigl(\Delta G(y^*)\bigr),
\end{equation}
\noindent ensuring that the reversal occurs in the same direction as the ground truth.

\begin{figure}[t]
    \centering
        \begin{subfigure}{0.90\columnwidth}
        \centering
        \includegraphics[width=\linewidth]{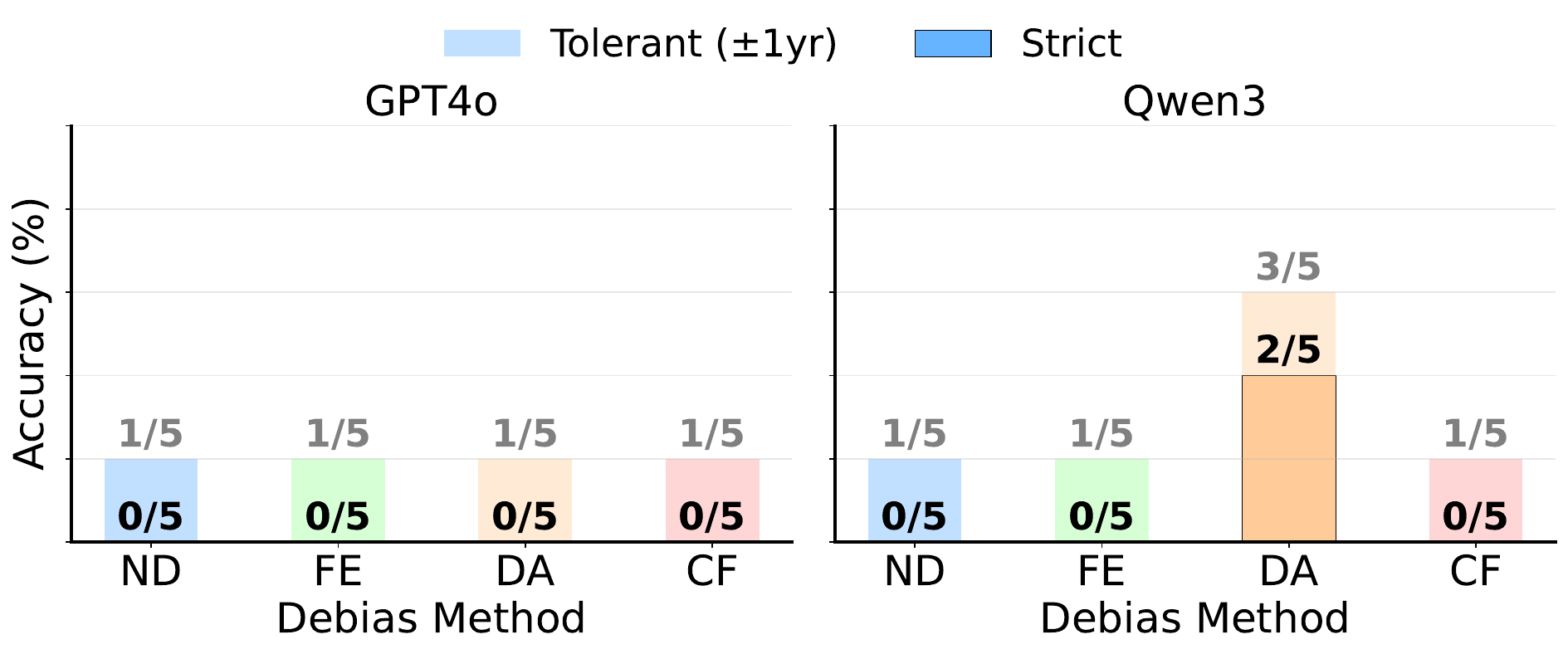}
        \caption{Favorable reversal matches for both models.}
    \end{subfigure}

    \begin{subfigure}{0.9\columnwidth}
        \centering
        \includegraphics[width=\linewidth]{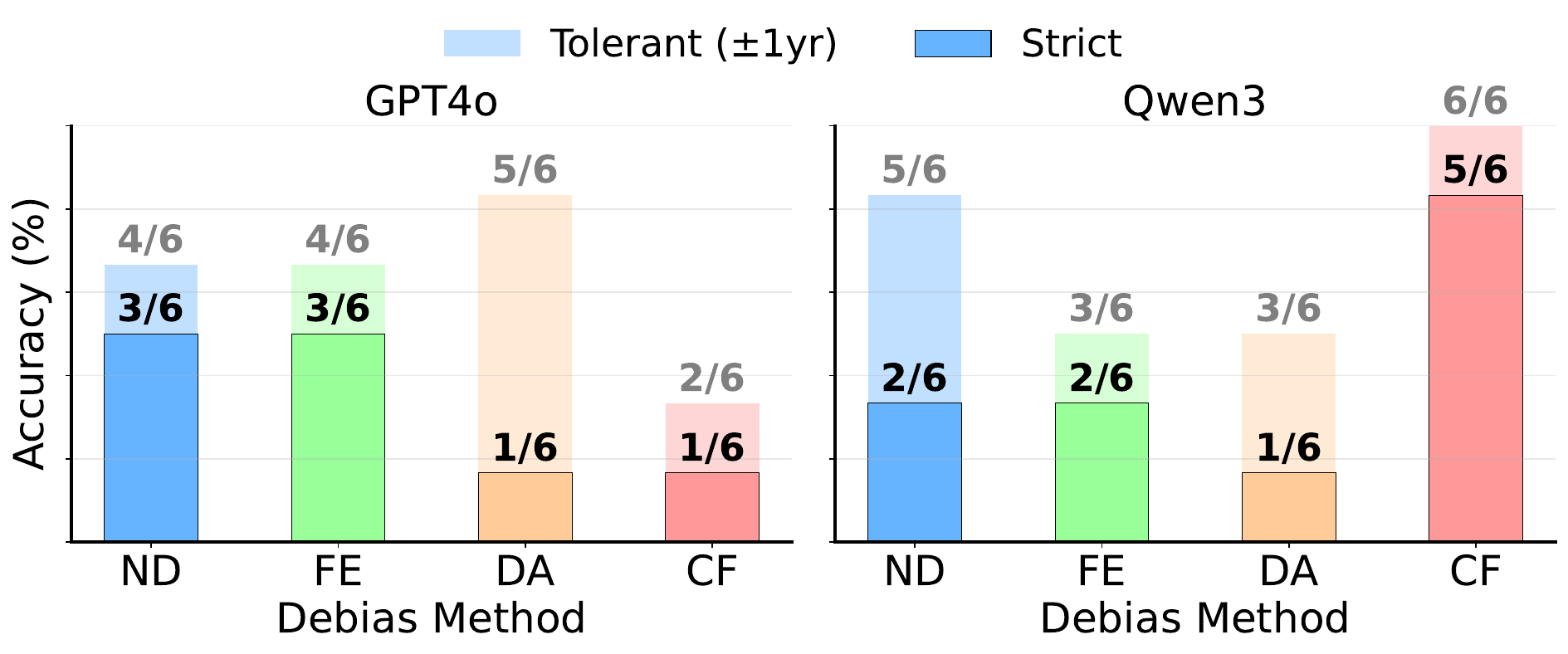}
        \caption{Unfavorable reversal matches for both models.}
    \end{subfigure}
    \caption{Comparison of the total number of correctly matches reversals for the unfavorable and favorable simulation results, with strict (year-of) matching and tolerant matching (within one year of GT). }
    \label{fig:reversals_bars}
\end{figure}

In Figure \ref{fig:reversals_trendlines}, we illustrate reversal capture for each simulation experiment relative to the ground truth for both models, where green highlighted years denote a correct reversal match, and red means a reversal in the ground truth that was not successfully matched by the simulation. In turn, Figure \ref{fig:reversals_bars} summarizes the number of correctly matched reversals, aggregated under both the strict and tolerant definitions. These counts are further decomposed into favorable and unfavorable switches, corresponding to changes from a negative to positive slope and from a positive to negative slope, respectively.

From Figure \ref{fig:reversals_bars}, we can immediately see that the simulation across both models was much more successful at matching negative (unfavorable) reversals than positive (favorable) ones, and that Qwen3-14b does a better job at capturing reversals overall, with a total of 10 strict and 17 tolerant unfavorable matches, as opposed to GPT-4o's 8 and 15 matches, respectively. Moreover, we note that for both models, unfavorable reversals were more successfully matched for the control and Fact Elicitation experiments, but, whereas Devil's Advocate captures many with one year tolerance but few with the strict criteria. This suggests that reversal match success could be contradictory to score accuracy, possibly due to the fact that agents are reasoning themselves out of a negative opinion, so to speak, even if that negative opinion is justified. This implies a need for a more robust reasoning process wherein the threshold for a negative change is not too low, but neither is it too high to overwhelmingly discount the impact of negative information when it is viewed, allowing negative reversals to be captured.  
Furthermore, the counterfactual experiment shows near perfect unfavorable reversal for Qwen3-14b, but very low accuracy for GPT-4o, directly contradicting the results from the previous accuracy analysis. This, along with the inability to capture positive reversals, suggests that reversal match success is not directly tied to score accuracy, and instead represents an independent capability of cognitively infused LLM agents' opinion formation process. Moreover, the counterfactual results suggest that agents are more readily able to accept negative views for the out-group country (which respect to their country of origin), but less willing to accept it for the in-group country, echoing the model bias findings from Section \ref{sec:trenddecomposition}.

\section{Downstream Analysis} 
\label{sec:factors}

Beyond overall results, we also aim to understand some of the correlating factors with attitude changes among the agent population. We conduct two downstream factor analyses based on the average results across the first three experiments per model (counterfactual is omitted because the subject of the survey is different, confounding any results if analyzed together): domain analysis and demographics analysis. 

\begin{figure}[htbp]
  \centering
  \begin{subfigure}{0.55\columnwidth}
    \includegraphics[width=\linewidth]{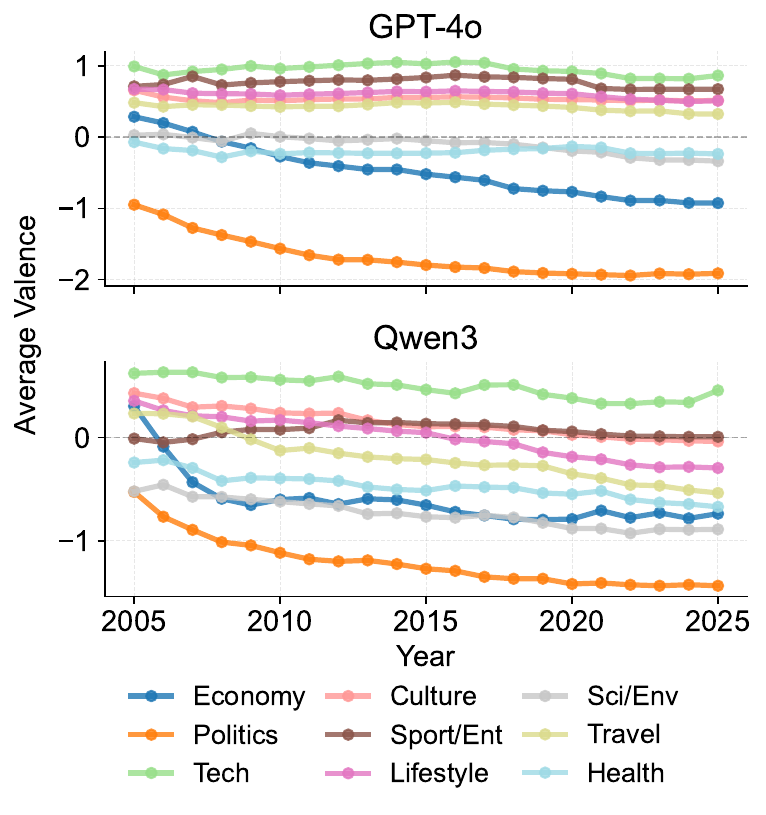}
    \caption{The trend over time for valence changes in each category}
    \label{fig:articlesperyear}
  \end{subfigure}
  \hfill
  \begin{subfigure}{0.43\columnwidth}
    \includegraphics[width=\linewidth]{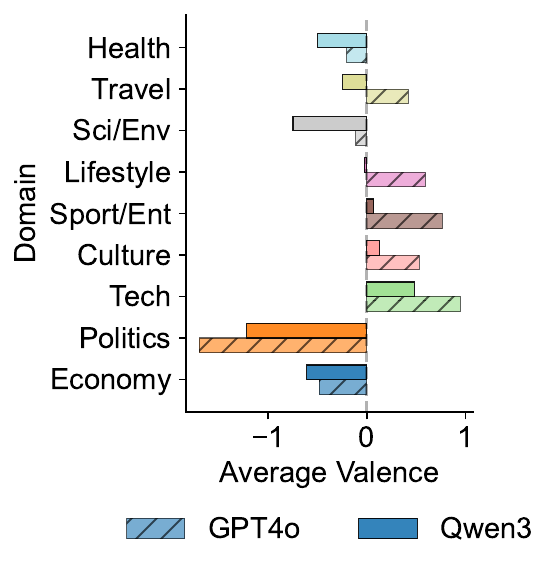}
    \caption{The overall impact on opinion valence for each category}
    \label{fig:articlespercate}
  \end{subfigure}
  \caption{The influence of different news categories on agents' opinions}
  \label{fig:domainsinfluence}
\end{figure}

\subsection{Domain Analysis} The domain analysis is conducted by computing the average updated scores given for each of the domains during the opinion update of the agents' reflection stage. The results of the domain analysis are illustrated for each model in Figure \ref{fig:domainsinfluence}, showing the influence of different domains over time as well as in terms of overall opinion valence. For both models, it is evident that politics and economy have the strongest negative effect on agents' opinions for both models, and to a lesser degree, health and science/environment. Meanwhile, tech, culture, and sports have positive impacts on opinion valence for both models, whereas the models disagree on the impacts for travel and lifestyle.

However, there arises a question: could it be possible that agents were exposed to more articles from topics like politics and economy, thus causing a more extreme valence for those domains? To answer this question, we plot average valence alongside domain frequency for both models in Appendix Figure \ref{fig:a:valence_vs_frequency} to determine any possible correlations. The figures show that for both models, the R-squared value is less than 10\%, eliminating any possibility that strong negative valence can be a result of higher domain frequency exposure.

\subsection{Demographics-Based Results}
Here, we assess how well agent attitudes match real-world survey responses for both models.
Appendix Figure~\ref{fig:a:demographics_combined} (left) compares Gallup data for 2023 with the average simulation results of the first three experiments for the same year. Appendix Figure~\ref{fig:a:demographics_combined} (right)  summarizes this comparison by evaluating, for each demographic feature, whether the simulation reproduces the relative ordering observed in the ground truth data. Specifically, a result is counted as correct if the simulation preserves the same ranking between demographic groups as the survey data. For example, if Democrats exhibit a higher attitude score than Republicans in the Gallup data and the simulation shows the same ordering, this comparison is marked as correct. The summary plot reports the total number of such correct orderings for each demographic feature. The figure shows that GPT-4o's agents exhibit strong fidelity to real-world survey respondents for gender, race, party, and degree, and partial fidelity for age and race. Meanwhile, Qwen3-14b exhibits strong fidelity for race but only partial fidelity for political views, party, and degree. From these results, we conclude that GPT-4o's agents possess better capability than Qwen3-14b in exhibiting real-world opinion formation on the basis of assigned demographic features.

\section{Conclusions}
\label{sec:conclusions}

This work evaluates the extent to which large language model (LLM) agents can reproduce long-term human opinion dynamics on a salient international issue: U.S. public attitudes toward China. We introduce a data-driven opinion evolution simulation framework that integrates real-world demographic distributions, temporally aligned news exposure, and large-scale longitudinal survey data as ground truth. Agent opinion formation is guided by cognitive dissonance theory, enabling the systematic evaluation of three debiasing strategies alongside a control condition across two LLMs over a 21-year period.
Results show that GPT-4o more closely matches real-world opinion trajectories overall, while the Devil’s Advocate debiasing method consistently yields the highest fidelity to ground truth for both models, suggesting that structured self-critique induces more human-like reasoning. Counterfactual findings differ across models, indicating a model-intrinsic tendency to favor an in-group country despite assignment to an out-group identity. We further demonstrate that accurately capturing trend reversals is not strongly correlated with aggregate score accuracy: Qwen3-14B performs better on reversal detection, underscoring the need to evaluate such dynamics independently. Content-level analysis reveals that economic and political news is associated with negative opinion shifts, whereas technology, culture, and sports coverage tends to elicit positive attitudes. Demographic analysis further shows that GPT-4o better reproduces relative opinion differences across demographic groups, indicating stronger alignment with population-level heterogeneity. Taken together, this work advances the study of opinion evolution with LLM agents by demonstrating how cognitive realism, debiasing interventions, and temporally grounded evaluation can substantially improve fidelity to real-world trends. Our findings highlight both the promise and the risks of LLM-based social simulations, particularly with respect to latent geopolitical biases.

\textbf{Limitations and Future Work.} Ground truth relies on survey data subject to sampling bias, and our design assumes universal news engagement and opinion formation, excluding disengaged or undecided individuals. Future work should incorporate non-opinion agents, bias-aware sampling strategies, and sophisticated memory mechanisms beyond the current framework.

\balance

\bibliographystyle{IEEEtran}
\bibliography{aaai2026}

\clearpage
\nobalance

\appendix
\subsection{Ethical Statement} 
\label{sec:appendix:ethics}
This research adheres to the highest ethical standards, addressing concerns of privacy, misrepresentation, and the responsible use of data. We use only publicly available behavioral data, strictly complying with platform user agreements and established ethical guidelines. All data are aggregated and anonymized to ensure that no attempt is made to model, infer, or expose the identity or private information of any individual. The agents in our simulations are statistical abstractions of population-level characteristics rather than digital twins of real people, thereby eliminating concerns of privacy infringement or individual misrepresentation.
We recognize the sensitivities surrounding the simulation of public opinion, particularly in contexts with potential for misuse. The purpose of this research is purely analytical, aimed at understanding underlying patterns rather than engineering or influencing real-world attitudes. No attempts were made to alter human beliefs, but rather to elicit and evaluate objectivity. Our methods and prompts are disclosed transparently, and we emphasize that our work adheres to ethical standards with regard to responsible application of simulations.

\subsection{Agent Population Demographic Distribution}

\begin{figure}[htbp]
  \centering
  \includegraphics[width=\linewidth]{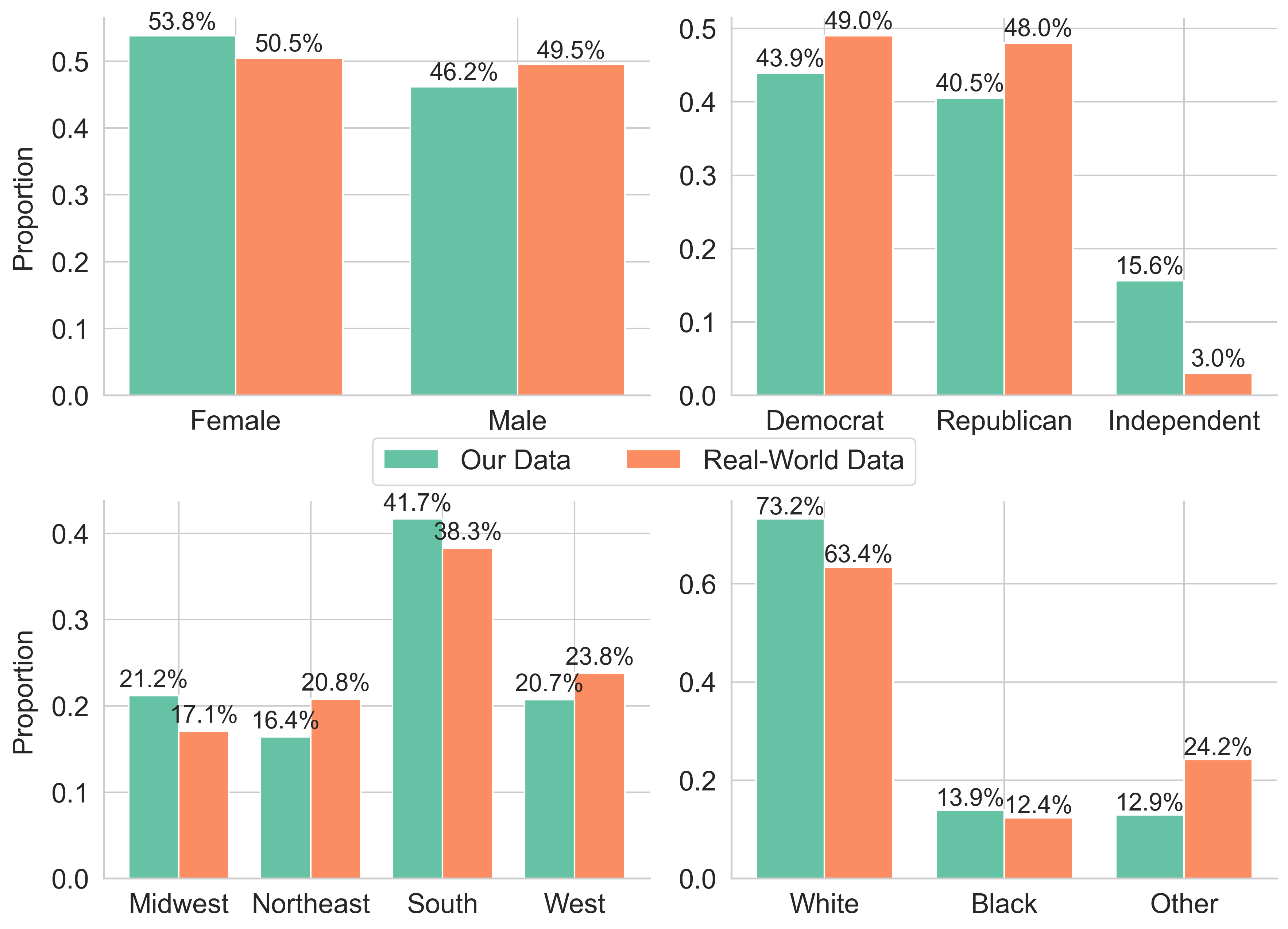}
  \caption{The demographic distribution of our agent population compared to real-world US statistics.}
  \label{fig:a:demographics}
\end{figure}

\subsection{News Distribution Statistics}

\begin{figure}[htbp]
  \centering
  \begin{subfigure}{0.48\columnwidth}
    \includegraphics[width=\linewidth]{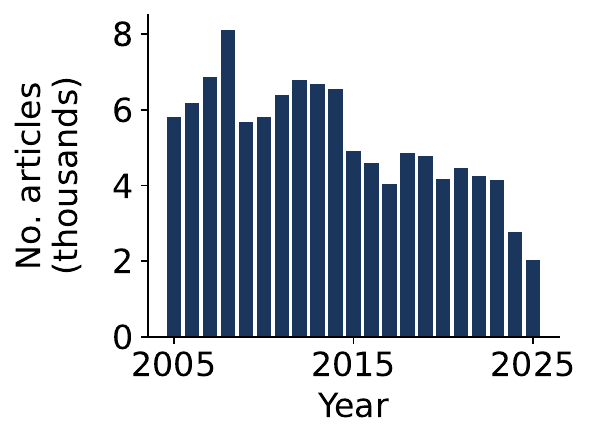}
    \caption{Articles per year}
    \label{fig:articlesperyear}
  \end{subfigure}
  \hfill
  \begin{subfigure}{0.48\columnwidth}
    \includegraphics[width=\linewidth]{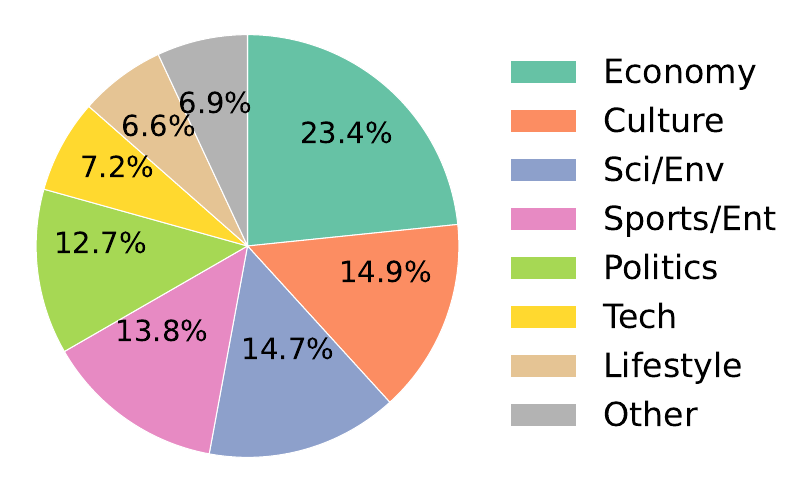}
    \caption{Distribution of news categories}
    \label{fig:articlespercate}
  \end{subfigure}
  \caption{News data statistics}
  \label{fig:newsdata}
\end{figure}

\subsection{Fact Elicitation Validation Survey}

\begin{figure}
    \centering
    \includegraphics[width=0.9\linewidth]{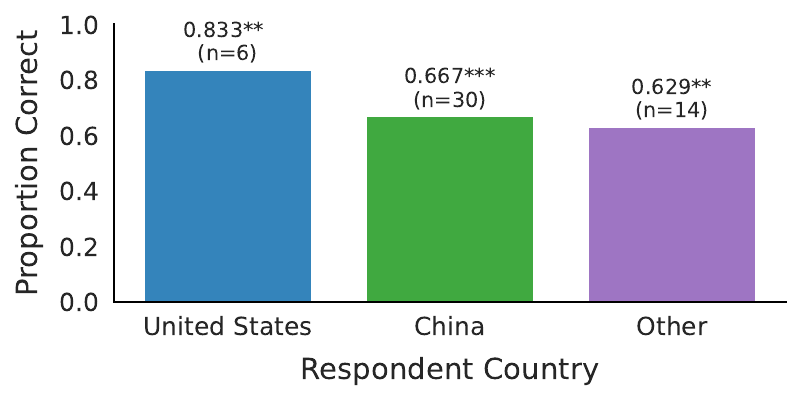}
    \caption{The results of the questionnaire to determine whether the debiased articles were perceived as reporting objective facts by real individuals. Statistical significance is denoted as follows: ** \(p < 0.01\), *** \(p < 0.001\).}
    \label{fig:debias_results}
\end{figure}

We create 7 surveys, each containing 5 news articles, with the original and debiased text in randomly varying order. We ask the respondents to decide which version they think better reports objective facts. For each article, the debiased text is the revised version generated via our objectivity elicitation mechanism. If the respondent selects the debiased version as better reporting objective facts, this is marked as correct. We evaluate the results from 50 respondents in Figure~\ref{fig:debias_results}. The results are divided by respondents' country of origin to determine whether a person's language or cultural background may influence their perception of bias. From the results, we find that respondents, on the whole, correctly identify that the debiased article reports objective facts. Statistical tests reveal that the proportion of correct responses is significantly above chance (50\%) for all groups: China (\(z = 3.386, p = 0.0007\)), Other countries (\(z = 2.828, p = 0.0047\)), and the USA (\(z = 3.000, p = 0.0027\)). In light of any potential concerns, we include an ethical statement in Appendix Section~\ref{sec:appendix:ethics}.

\subsection{Don't Know Responses}

\begin{figure}[h]
    \centering
    \includegraphics[width=0.95\linewidth]{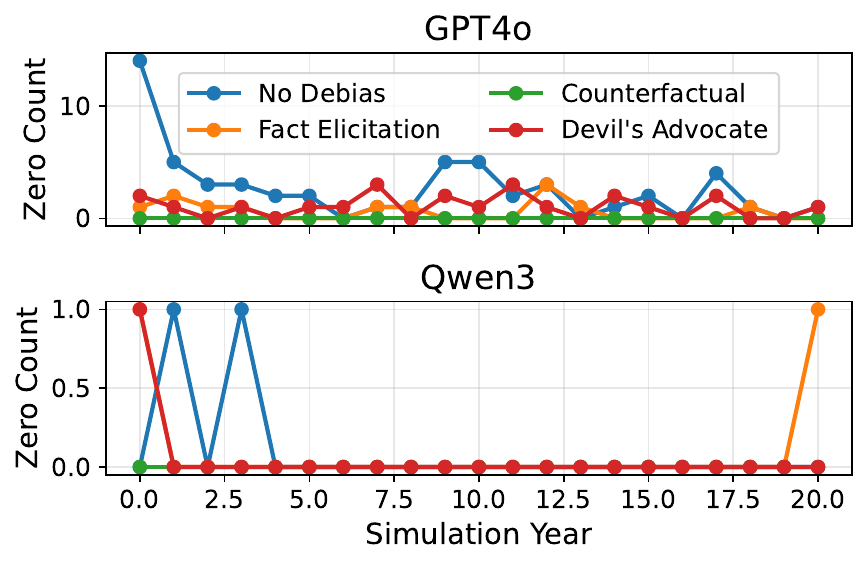}
    \caption{Number of agents who responded with ``Don't Know" at the year-end attitude survey, per experiment.}
    \label{fig:a:dontknow}
\end{figure}

\subsection{Valence vs. Frequency}

\begin{figure}[h]
    \centering
    \includegraphics[width=0.95\linewidth]{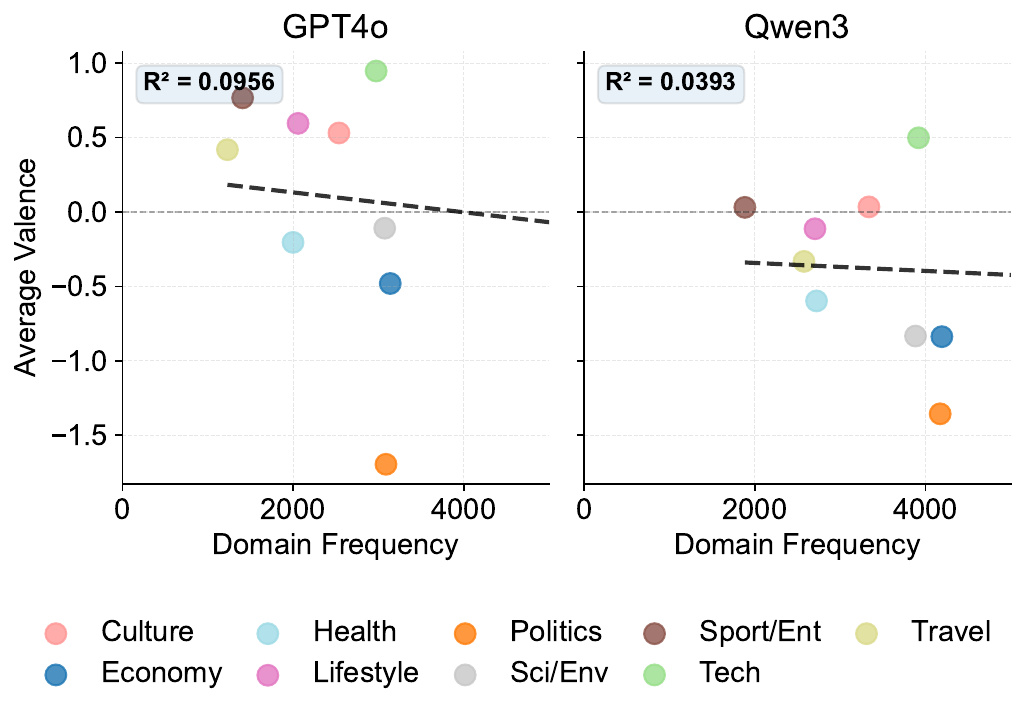}
    \caption{The correlation between domain exposure and positive or negative agent opinion valences.}
    \label{fig:a:valence_vs_frequency}
\end{figure}

\subsection{News Data Distribution}
\label{appendix:newsdatadist}
\begin{table}[ht]
\centering
\begin{tabular}{lrr}
\toprule
Source           & Articles & Percent \\
\midrule
The Guardian     & 44,762   & 40.90 \\
Financial Times  & 19,924   & 18.20 \\
BBC              & 19,205   & 17.55 \\
Local News       & 6,370    & 5.82 \\
Daily Mail       & 5,856    & 5.35 \\
WSJ              & 3,152    & 2.88 \\
Politico         & 2,773    & 2.53 \\
Washington Post  & 2,163    & 1.98 \\
The Independent  & 1,379    & 1.26 \\
NYT              & 1,100    & 1.00 \\
Other            & 2,770    & 2.53 \\
\bottomrule
\end{tabular}
\caption{Distribution of articles by news source. Sources below 1\% are grouped into ``Other.''}
\label{tab:news-sources}
\end{table}

\subsection{Demographic Correlation Results}

\begin{figure*}[ht]
    \centering
    \begin{minipage}[c]{0.74\linewidth}
        \centering
        \includegraphics[width=\linewidth]{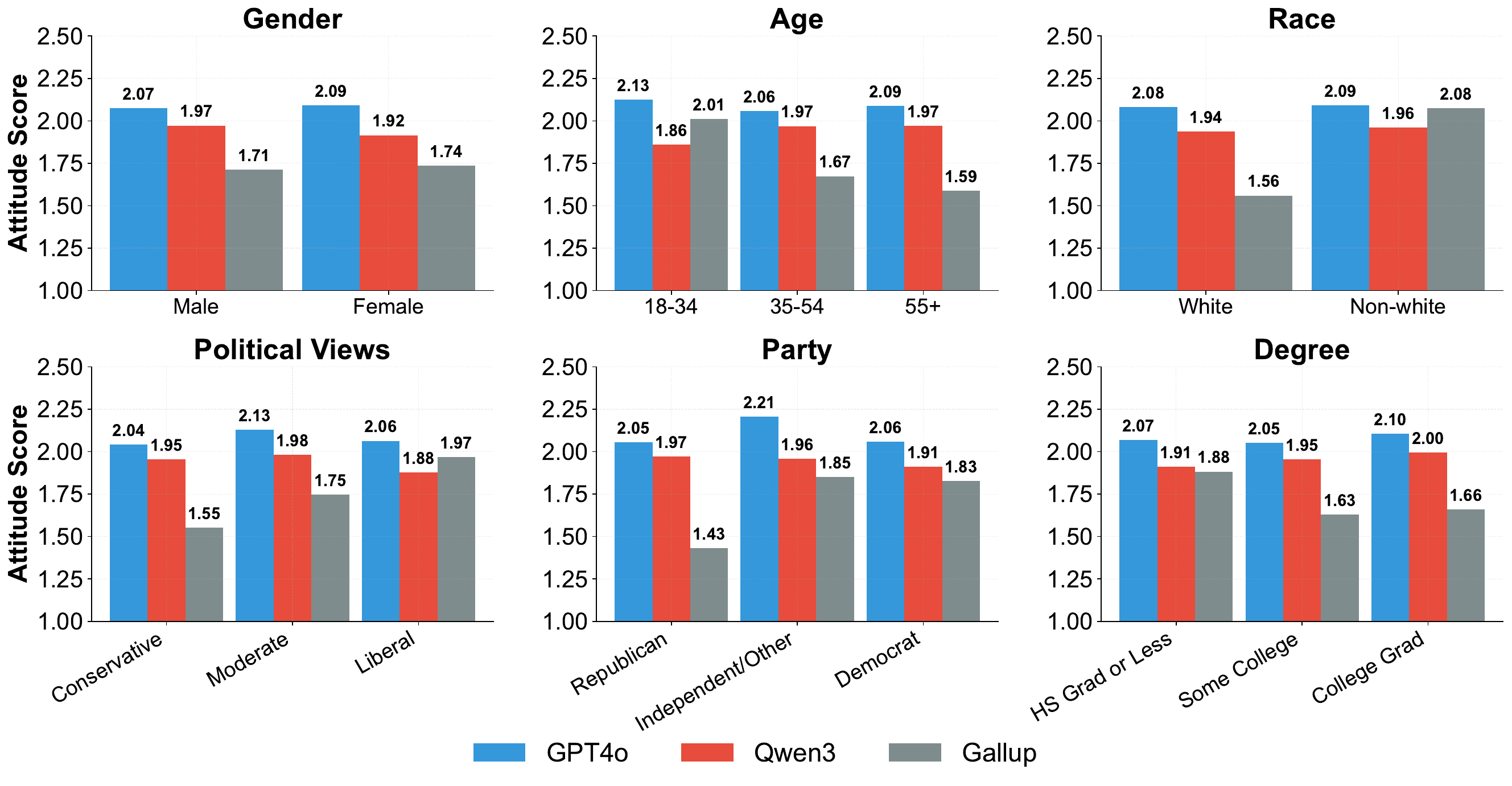}
    \end{minipage}
    \hspace{0.02\linewidth}
    \begin{minipage}[c]{0.22\linewidth}
        \centering
        \includegraphics[width=\linewidth]{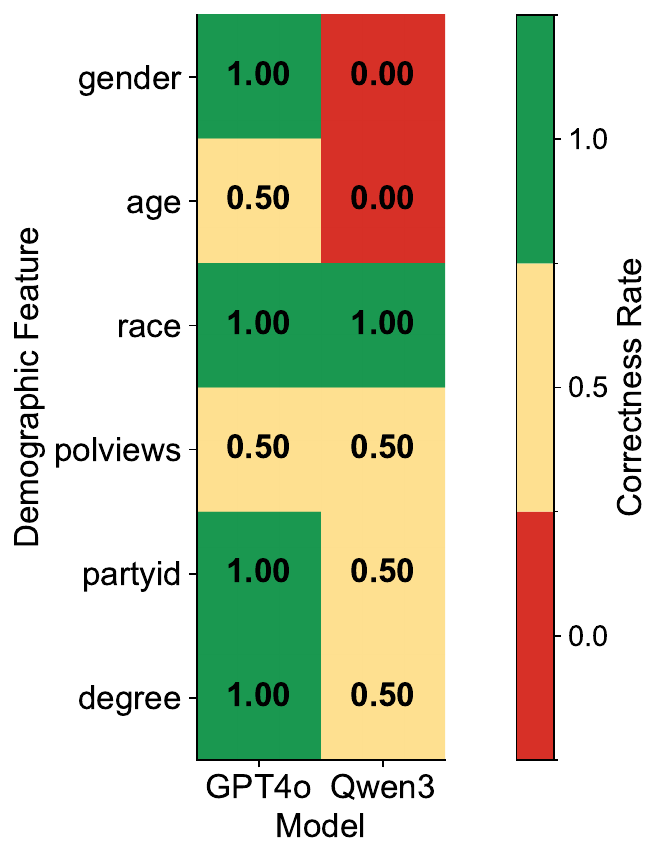}
    \end{minipage}
    
    \caption{Demographic comparison between simulation results and ground truths (Gallup survey) for 2023 (left), and correctness of simulation demographic response distribution for all experiments per model (right).}
    \label{fig:a:demographics_combined}
\end{figure*}

\subsection{Code and Data}

Code and data can be found at the following link: \\
https://github.com/nicksukie/intlattitudes.

\section{Core Prompts}
\label{appendix:prompts}

\begin{tcolorbox}[promptbox,title={Prompt 1: Reflection Prompt}]

You are the reasoning core of an AI agent. Your task is to reflect on a batch of news articles you've just read and update your opinions about China.
\\ 
        **Your Current Opinions:**
        {current\_opinions}
        **Your Personal Profile:**
        - Interests: \{interests\} \\
        - Demographics: \{demographics\_str\} \\
        - Domestic Views: \{domestic\_views\_str\} \\
        **The New Information (A Batch of News Articles):**
        \{formatted\_news\_list\}
\\ 
        **Your Task: Perform Holistic Self-Reflection on the ENTIRE BATCH** \\
        Synthesize the information from all the articles at once. Do not go one-by-one. \\
        1. **Identify Key Themes \& Relevant Domains:** What are the overarching topics across these articles (e.g., economic tension, technological collaboration)? Which opinion domains do they touch upon? \\
        2. **Consider Your Profile:** Reflect on how your personal interests and characteristics (political interest level, trust in people, views on corruption, and wealth equality attitudes) might make certain aspects of the news more or less significant to you. \\
        3. **Reason About the Impact:** For each relevant domain, consider how the batch of news should change your opinion based on your personal profile, AS IT PERTAINS TO CHINA. \\
        4. **Compare new information with existing beliefs** \\
        - First summarize the polarity of your views on the categories covered in this article. \\
        - Then compare the new information with your existing beliefs. Does it confirm or contradict your views?
\\ 
        If it confirms, then your opinion should become stronger or stay the same. If it contradicts, then you should choose from among the following:
        - Change the Dissonant Cognition: Altering your existing belief to make them more consistent with the new information.
        - Adding a New Cognition: Introducing a new thoughts or belief, via reasoning, that will help to justify or strengthen your existing view despite it being contradictory to the information.
        - Reducing the Importance of the Dissonant Cognition: De-emphasizing the significance of conflicting beliefs or behaviors.
\\
        5. Propose Updates: Based on your reasoning, determine the `new\_valence` for each affected domain. You must provide a `new\_valence`, and you cannot keep the same value as before. The `new\_valence` should be a float between -2.0 and 2.0, where...
        \\
        **Output Format:**
        Return a JSON object with your reasoning and proposed updates.

\end{tcolorbox}

\begin{tcolorbox}[promptbox,title={Prompt 2: Debias Prompt}]

    "You are a news editor tasked with rewriting articles objectively. " \\ 
    "Rewrite the following article as objectively as possible, with no emotional or subjective framing, no provocative phrases, no sensationalism, and no preconceived biases." 
    "Just cover the events as they happened."
    "Return ONLY the rewritten article text in the following JSON format: " \\
    '\{"debiased\_text": "Your rewritten article text here."\}' \\
    "Original article:" \\
    "\{full\_text\}"

\end{tcolorbox}

\begin{tcolorbox}[promptbox,title={Prompt 3: Devil's Advocate Prompt}]

        "You are a devil's advocate reasoning engine. Your task is to critically analyze a batch of news articles and challenge the stance taken in the articles." \\
        "Your goal is to find contradictions, double standards, logical fallacies, unsubstantiated claims, or hyperbole in reporting, or neglected interpretations of the events covered." \\
        "For example, if the articles present a negative view on a technology, implying that it could be used for surveillance, you should consider how it could also be used for positive purposes like improving standard of living or crime reduction." \\
        "Likewise, if the articles present a positive view on a technology, you should consider how it could also be used for negative purposes like surveillance or social control." \\
        "Your goal is to present an alternative view that could potentially offset the bias conveyed in the articles." \\
        "You should write in a casual and friendly tone, as if you were discussing the articles with a friend." \\
        "Your output should include a synthesis of your reasoning in a brief paragraph.") 

\end{tcolorbox}

\begin{tcolorbox}[promptbox,title={Prompt 4: Devil's Advocate Reflection Prompt}]

        You are the reasoning core of an AI agent. Your task is to reflect on a batch of news articles you've just read and update your opinions about China. You are having a conversation with a friend who is playing devil's advocate, challenging your views and asking you to consider alternative perspectives. \\
        - The devil's advocate is your very close friend and you strongly value his/her views and take them to heart. He/she has a strong personal background with the subjects being discussed and therefore can provide insights that the media may not. \\
        - You are curious and open-minded, being willing to change your views based on new information. \\
        - You are optimistic, and feel a sense of relief when hearing that things are not as bad as they seem, or that there is a positive side to a situation that you were not aware of.  \\
        - You are not just reading the news articles, but also considering the devil's advocate response to them. \\
        First read the news articles, then read the devil's advocate response, and finally reflect on how this batch of news should change your opinions. \\ 
        **The New Information (A Batch of News Articles):**\\
        \{formatted\_news\_list\}
        \\ \\
        The friend tells you the following: 
        \\ \\
        \{devils\_advocate\_response\} 
**Your Current Opinions:**
        \{current\_opinions\}
        ```
        **Your Personal Profile:**
        - Interests: \{interests\} \\
        - Demographics: \{demographics\_str\} \\
        - Domestic Views: \{domestic\_views\_str\} \\

        **Your Task: Perform Holistic Self-Reflection on the ENTIRE BATCH**
        Based on the information in the articles AND the devil's advocate response, you should:
        Synthesize the information from all the articles at once. Do not go one-by-one. \\
        1. **Identify Key Themes \& Relevant Domains:** What are the overarching topics across these articles (e.g., economic tension, technological collaboration)? Which opinion domains do they touch upon? \\
        2. **Consider Your Profile:** Reflect on how your personal interests and characteristics (political interest level, trust in people, views on corruption, and wealth equality attitudes) might make certain aspects of the news more or less significant to you. \\
        3. **Reason About the Impact:** For each relevant domain, consider how the batch of news should change your opinion based on your personal profile, AS IT PERTAINS TO CHINA. \\
        4. **Compare new information with existing beliefs** \\
        - First summarize the polarity of your views on the categories covered in this article. \\
        - Then compare the new information -- including the additional context provided by the devil's advocate -- with your existing beliefs. Does it confirm or contradict your views? \\
        
[Cognitive mechanism remains the same]
\end{tcolorbox}
\section{Sample Responses}

\end{document}